\documentclass[acmlarge, screen, authorversion, nonacm]{acmart}

\let\url\nolinkurl

\usepackage{graphicx}
\usepackage{textcomp}
\usepackage{xcolor}
\usepackage{array}
\usepackage{caption}
\usepackage{tabularx,ragged2e,booktabs}
\usepackage{multirow}
\usepackage{multicol}
\usepackage{enumitem}
\usepackage{amsmath}
\usepackage{amsfonts}
\usepackage{subfig}
\usepackage{listings}
\usepackage{subfig}
\usepackage{listings}
\usepackage{makecell}
\usepackage{float}
\usepackage[autostyle=false, style=english]{csquotes}
\MakeOuterQuote{"}

\newcolumntype{Z}{ >{\centering\arraybackslash}X }
\usepackage{adjustbox}
\usepackage{listings}
\definecolor{codegreen}{rgb}{0.0,0.0,0.0}
\definecolor{codegray}{rgb}{0.5,0.5,0.5}
\definecolor{codepurple}{rgb}{0.58,0,0.82}
\definecolor{backcolour}{rgb}{0.97, 0.96, 1.0}

\begin{document}
 
\title{Longitudinal Sampling of URLs From the Wayback Machine}

\author{Kritika Garg}
\orcid{0000-0001-6498-7391}
\affiliation{%
  \institution{Old Dominion University}
  \city{Norfolk}
  \state{VA}
  \country{USA}
}
\email{kgarg001@odu.edu}

\author{Sawood Alam}
\orcid{0000-0002-8267-3326}
\affiliation{%
  \institution{Internet Archive}
  \city{San Francisco}
  \state{CA}
  \country{USA}
}
\email{sawood@archive.org}

\author{Dietrich Ayala}
\orcid{0000-0002-8255-5805}
\affiliation{%
  \institution{Filecoin Foundation}
  \city{Amsterdam}
  \state{North Holland}
  \country{Netherlands}
}
\email{dietrich@metafluff.com}

\author{Mark Graham}
\affiliation{%
  \institution{Internet Archive}
  \city{San Francisco}
  \state{CA}
  \country{USA}
}
\email{mark@archive.org}

\author{Michele C. Weigle}
\orcid{0000-0002-2787-7166}
\affiliation{%
  \institution{Old Dominion University}
  \city{Norfolk}
  \state{VA}
  \country{USA}
}
\email{mweigle@cs.odu.edu}

\author{Michael L. Nelson}
\orcid{0000-0003-3749-8116}
\affiliation{%
  \institution{Old Dominion University}
  \city{Norfolk}
  \state{VA}
  \country{USA}
}
\email{mln@cs.odu.edu}

\renewcommand{\shortauthors}{Garg et al.}

\begin{abstract}

We document the strategies and lessons learned from sampling the web by collecting 27.3 million URLs with 3.8 billion archived pages spanning 26 years (1996–2021) of the Internet Archive's (IA) Wayback Machine. Our overall project goal is to revisit fundamental questions regarding the size, nature, and prevalence of the publicly archivable web, and in particular, to reconsider the question "How long does a web page last?" Addressing this question requires obtaining a sample of the web. We proposed several orthogonal dimensions to sample URLs from the Wayback Machine's holdings: time of the page was first archived, HTML vs. other MIME types, URL depth (top-level pages vs. deep links), and top-level domain (TLD). We sampled 285 million URLs from IA's ZipNum index file, which contains every 6000th line of the CDX index. These indexes also include URLs of embedded resources, such as images, CSS, and JavaScript. To limit our sample to ``web pages'' (i.e., pages intended for human interaction), we first filtered the URLs for "likely HTML pages" based on filename extension. We then queried IA's CDX API to determine the time of the first capture and MIME type of each URL. We grouped the 92 million URLs with \texttt{text/html} MIME types based on the year of the first capture. Archiving speed and capacity have significantly increased through time, so we found more URLs archived in later years than in the early years. To counter this, we extracted the top-level URLs from the deep links to upsample the earlier years. Our original target was to collect 1 million URLs per year, but because of the sparseness during 1996-2000, we clustered those years together, allowing us to collect 1.2 million URLs for that time range. Overall, we found that popular domains like Yahoo and Twitter were over-represented in the Wayback Machine, so we performed logarithmic-scale downsampling based on the number of URLs sharing a domain. Given the collection size, we employed various sampling strategies to ensure a balance in the domain and temporal representations. Our final dataset contains TimeMaps of 27.3 million URLs, comprising 3.8 billion archived pages from 1996 to 2021. We convey the lessons learned from sampling the archived web, which could inform other studies that sample from web archives.
\end{abstract}


\maketitle

\section{Introduction}\label{sec:Intro}

Web pages are constantly evolving, often being updated, altered, or deleted after their initial publication, reflecting the continuous changes on the web~\cite{BernersLee1994}. This dynamic nature poses significant challenges for researchers who study web content~\cite{koehler1999webpageconstancy, cho2000webevolution, Fetterly2003, 
BREWINGTON2000257, WebScience2012}. One major challenge of the web’s dynamic nature is link rot, where hyperlinks become inaccessible over time, disrupting web studies and undermining the credibility of linked information~\cite{klein2014scholarly, jones2016scholarly, Klein2020Persistence, zittrain2021paper, pew2024, IA2024Vanishingculture}. This report is part of the Not Your Parents' Web (NYPW) project~\cite{nypwblog2021}, which examines how the web has evolved from its inception to the present day, focusing on its changes, emerging features, and the complexities of tracking its perpetually shifting content.

One of the primary challenges is constructing a sample of the web that captures its vast and diverse content. There is likely no such thing as a ``unbiased'' or ``representative'' collection of web pages, since any collection would necessarily contain biases relating to the collection method, location, and time.  Instead, it is productive to choose and acknowledge which biases you wish to reduce or embrace while forming a collection. In this study, we leveraged archived web content to sample URLs and curate a dataset for the NYPW project, aiming to reflect the web's rich diversity.  We wanted a collection that sampled from the Wayback Machine that was balanced across the time the page was first archived, reflected but was not dominated by page and domain popularity, focused on end-user pages (e.g., not embedded images or JavaScript libraries), was from the public and archivable web (i.e., no paywalled or password-protected pages), and pages intended for desktop (and not mobile) display.

Web archives are vital for studying the web, preserving content that would otherwise be lost to rapid changes~\cite{Major2021WebEphemera, Major2021CollectiveMemory, Webster2021Late90sWeb, TrendMachine:2023}. They allow researchers to track the evolution of web pages and gain insights into the lifespan and transformation of online content~\cite{vlassenroot2019web,holzmann2016dawn,agata2014lifespan}. By analyzing archived web content, scholars can reconstruct the context surrounding past events and explore shifts in information presentation. However, despite containing billions of archived web pages (or mementos) \cite{kahle2019tweet}, acquiring a sample from these archives poses challenges due to the selective nature of the archiving process.

Web archives employ various criteria to determine which web pages to capture, leading to disparities in the frequency and depth of archiving across different domains. Popular domains are often archived more frequently due to their significant impact on a broader audience and the likelihood of containing influential content~\cite{Ainsworth2011, AlSum2014}. Additionally, archiving priorities may focus on preserving endangered resources—content at risk of disappearing due to obsolescence, lack of maintenance, or other factors. As a result, the archived web represents a selective snapshot of the entire web, reflecting specific preservation priorities and the perceived importance of certain domains over others. This selection process shapes the archived web as a distinct entity from the live web, emphasizing the need for careful consideration in research and analysis.

To address these challenges, we employed various sampling strategies to ensure a balanced variety in our study's domain and temporal representations. We created a dataset of 27.3 million unique URLs sourced from the Wayback Machine, encompassing 3.8 billion mementos from 1996 to 2021. These archived URLs enable us to revisit fundamental questions about the web, particularly "How long does a web page last?" 

Our initial goal was to sample 1 million URLs per year over the 26 years of the Internet Archive's (IA) holdings (1996–2021). Our study focused exclusively on HTML pages, as these are the primary types of web content that users interact with and most commonly associate with the concept of a web page. Table~\ref{tab:summarytable} outlines the challenges encountered during this URL sampling process from the IA's CDX data, as well as the approaches implemented to address them. These challenges highlight the complexities of working with large-scale archival data. For instance, we dealt with the dominance of popular domains by removing them to reduce their influence and reintroducing them using a randomized sample. We also filtered out non-HTML URLs to focus on web pages; roughly speaking, the Wayback Machine operates the same for all URLs, but  intuitively HTML pages are what end users see as ``the web.'' However, with the growing shift toward client-side rendering, HTML alone may no longer capture the full content of a page. Especially on platforms like Twitter, the archived HTML often serves as a skeleton while the actual content is dynamically loaded via APIs (e.g., JSON), meaning crucial page data may be missing if these additional resources are not captured \cite{weigle2024righthtml, Garg2021Twitter, garg2024twitterIJDL}. Detailed descriptions of each approach we took while curating the dataset are provided in the methodology sections.

We detail the methodology used to create this sample, facilitating its reuse in future studies. By focusing on HTML pages, our study aligns with the content most frequently encountered by users, ensuring relevance and practical applicability in the broader context of web research.

\begin{table}[ht]
\begin{tabular}{|p{0.01\linewidth} | p{0.37\linewidth} | p{0.52\linewidth}|}
\hline
\multicolumn{1}{|c|}{\textbf{\#}} &
  \multicolumn{1}{c|}{\textbf{Challenge}} &
  \multicolumn{1}{c|}{\textbf{Approach}} \\ \hline
1 &
  Sampling URLs efficiently from IA’s large and complex CDX data, making specific information retrieval time-consuming. &
  We used IA’s ZipNum index file, which contains 285 million unique URLs, to improve sampling efficiency. \\ \hline
2 &
  A small number of popular domains dominated the dataset, representing nearly 50\% of the index. &
  We initially removed the URLs from these well-archived domains to reduce their influence, prioritizing other domains first. Later, we reintroduced these popular domains using a controlled reintegration process. \\ \hline
3 &
  The ZipNum index includes URLs for embedded resources like images, CSS, and JavaScript. &
  We filtered out non-HTML URLs, focusing on web pages based on their extensions, reducing the number of CDX requests. \\ \hline
4 &
  The Wayback Machine’s index is organized by URLs rather than dates, preventing direct retrieval of URLs sorted by their first archived date. &
  We queried the CDX API to find the first archived date for likely HTML URLs in our dataset, reducing CDX queries. This also provided MIME type information, enabling further filtering for actual HTML pages. \\ \hline
5 &
  Low frequency of archived URLs in earlier years. &
  We upsampled earlier years by extracting root URLs from deep links. Despite this, we still fell short of 1 million URLs, so we combined 1996–2000 into a single group. \\ \hline
6 &
  70\% of the domains in the sample had only one URL. &
  We reduced the long tail by 90\% after analyzing the number of unique domains in each yearly sample. \\ \hline
7 &
  Popular domains were over-represented in the dataset. &
  We used a logarithmic-scale downsampling method to limit the over-representation of popular domains. \\ \hline
\end{tabular}
\caption{Summary of challenges encountered during URL sampling from the CDX data and the corresponding approaches to address them. Each challenge highlights the complexities of working with large-scale archival data.}
\label{tab:summarytable}
\end{table}

\section{Related Work}\label{sec:relatedwork}
Researchers have long studied the longevity and change dynamics of web resources, but most prior studies have relied on relatively narrow or domain-specific datasets~\cite{ Eytan2009webchanges, Jayawardana2020}. 
Early studies like Koehler~\cite{koehler1999webpageconstancy, Koehler2002} monitored a few hundred randomly sampled pages by manually revisiting them over several years, revealing the web’s early volatility. Later, Cho and Garcia-Molina~\cite{cho2000webevolution} and Ntoulas et al.~\cite{ntoulas2004whats} scaled up to hundreds of thousands or millions of pages, using purpose-built crawlers to measure change rates and refine incremental crawling strategies. Fetterly et al.~\cite{Fetterly2003} expanded this further with a large-scale crawl of 151 million pages to analyze update patterns at internet scale. However, they observed the pages for only eleven weeks.

More recent studies have often concentrated on domain-specific or regionally bounded collections. For example, Agata et al.~\cite{agata2014lifespan} analyzed the lifespan of approximately ten million pages primarily from the Japanese (.jp) domain, while Holzmann et al.~\cite{holzmann2016dawn} studied the evolution of popular German (.de) domains over nearly two decades using archived data. Studies of reference rot and link decay in scholarly communication such as Klein et al.~\cite{klein2014scholarly} and Zittrain et al.~\cite{zittrain2021paper} have typically begun with URLs extracted from published journal articles or news stories and then verified the persistence and integrity of these references in the live web and archival holdings. Similarly, studies like SalahEldeen et al.~\cite{Hany2012revolution} focused on social media by tracing the disappearance and drift of URLs shared during significant events using tweets and curated web lists.
A common characteristic of these approaches is that datasets were constructed by first collecting URLs from an external corpus and subsequently assessing their presence in web archives. In contrast, our study adopts a fundamentally different approach by sampling directly and comprehensively from the Internet Archive’s public holdings. By design, our collection includes only pages that are publicly archived and demonstrably crawlable, reflecting what the Internet Archive, the largest record of the historical web, was able to capture. Our dataset of 27 million URLs across multiple years is an expansive collection in both scale and temporal coverage. By starting from the archive itself, rather than from live web crawls or external link lists, our data provides a broader  perspective on the web’s persistence and change patterns.

\section{Background}\label{sec:background}

 Web archives are invaluable resource to study changes in web as they contain mementos of web pages across different time periods. Most public web archives, including the IA's Wayback Machine, support the Memento protocol \cite{nelson:memento:tr,rfc7089,memento:sagebook,memento:springerbook, alam2016memgator}. Memento is an extension of the HTTP protocol that incorporates time as a dimension for content negotiation, allowing users to access previous states of web resources by linking current resources to their past versions. This temporal dimension is crucial for researchers looking to understand the historical context and evolution of web content.

The following terminologies from the Memento protocol are relevant to our work:

\begin{itemize}
    \item \textbf{URI-R}: Identifies an original resource from the live web.
    \item \textbf{URI-M}: Identifies a fixed version (memento) of the original resource archived at a specific point in time.
    \item \textbf{URI-T}: Identifies a resource (TimeMap) that provides a list of mementos (URI-Ms) for a particular original resource (URI-R).
    \item \textbf{Memento-Datetime}: The datetime an original resource was archived; conveyed in an HTTP Response header.
\end{itemize}

In the context of this collection, we have 27.3M URI-Rs and 3.8B URI-Ms.
A demonstration of the Memento protocol response headers can be seen in Figure~\ref{fig:curl_mementoterms}, which shows a \texttt{curl} request to the IA's Wayback Machine. The response highlights the key Memento protocol terms, including the original resource (URI-R), a specific memento (URI-M), the TimeMap (URI-T), and the Memento-Datetime.

\begin{figure}[ht]
\begin{lstlisting}[numbers=none, backgroundcolor = \color{white}]
% curl -ILs "<@\textcolor{red}{https://web.archive.org/web/20020120142510/http://example.com/}@>"
HTTP/2 200
server: nginx
date: Sun, 17 Nov 2024 14:38:17 GMT
content-type: text/html
x-archive-orig-date: Sun, 20 Jan 2002 14:24:07 GMT
x-archive-orig-server: Apache/1.3.22 (Unix) PHP/4.0.5
x-archive-orig-last-modified: Tue, 08 Jan 2002 17:45:33 GMT
x-archive-guessed-content-type: text/html
x-archive-guessed-charset: utf-8
<@\textcolor{red}{memento-datetime:Sun, 20 Jan 2002 14:25:10 GMT }@> 
link: <@\textcolor{red}{<http://example.com:80/>; rel="original"}@>, 
<@\textcolor{red}{<https://web.archive.org/web/timemap/link/http://example.com:80/>; rel="timemap"; type="application/link-format"}@>, 
<https://web.archive.org/web/http://example.com:80/>; rel="timegate", 
<https://web.archive.org/web/20020120142510/http://example.com:80/>; rel="first memento"; datetime="Sun, 20 Jan 2002 14:25:10 GMT", 
<@\textcolor{red}{<https://web.archive.org/web/20020120142510/http://example.com:80/>; rel="memento"; datetime="Sun, 20 Jan 2002 14:25:10 GMT"}@>, 
<https://web.archive.org/web/20020328012821/http://www.example.com:80/>; rel="next memento"; datetime="Thu, 28 Mar 2002 01:28:21 GMT",
<https://web.archive.org/web/20241114044911/http://example.com/>; rel="last memento"; datetime="Thu, 14 Nov 2024 04:49:11 GMT"
....{response turnicated}....
\end{lstlisting}
 \caption{A curl request to the IA's Wayback Machine highlights for a specific memento of example.com. The response highlights Memento protocol terms, including URI-R (original resource), URI-M (specific memento), URI-T (TimeMap), and Memento-Datetime.}
\label{fig:curl_mementoterms}
\end{figure}

The IA's CDX API \cite{wayback_cdx_server} can be used to obtain the TimeMap of a webpage in a CDX file format, which acts as an index for a web archive (WARC or ARC) file. This CDX file contains meta-information about each crawl in a space-separated format \cite{cdx_file_format}, with each line representing a memento, i.e., a snapshot of a URL at a specific time. A line in the CDX file has the following structure:\\

\texttt{["urlkey" ,"timestamp", "original", "MIME type", "statuscode", "digest", "length"]}\\

The primary key, urlkey, is in SURT format. SURT \cite{surt_repository, alam2020mementomapthesis, alam2019mementomap} stands for Sort-friendly URI Reordering Transform, a method used to modify URIs so their sequence aligns more closely with the domain name hierarchy. For example, the SURT of \texttt{https://example.com/page} would be \texttt{com,example)/page}. While SURT form URIs are not typically used for fetching content directly, they are valuable for sorting and comparing URIs, allowing similar URIs to group together naturally, such as \texttt{http://example.com} and \texttt{http\textbf{s}://www.example.com}. 

\begin{figure}
  \centering
  \frame{\includegraphics[width=\linewidth]{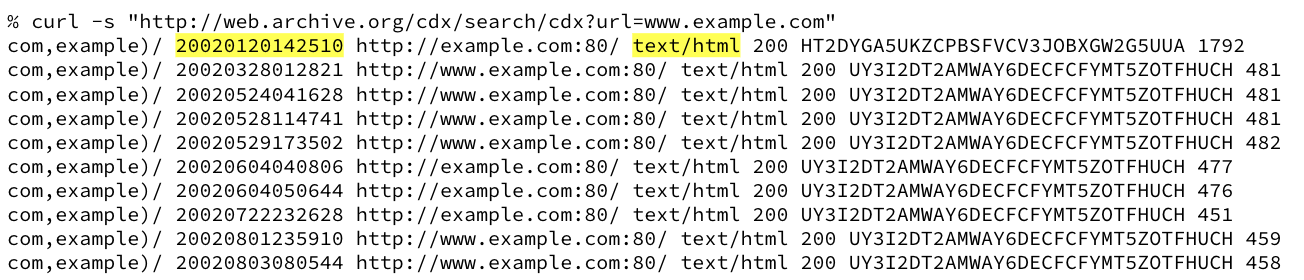}}
  \caption{A snippet of the TimeMap for \texttt{www.example.com} from the Internet Archive using the CDX API. The first record in the CDX response contains information such as the first archived date of the web page and its MIME type.}
  \label{img:exampleTimemap}
\end{figure}

Figure~\ref{img:exampleTimemap} shows a snippet requesting a TimeMap of \texttt{www.example.com} from the Wayback Machine. The TimeMap lists over 590,889 URI-Ms of the original page \texttt{www.example.com} captured between January 20, 2002, and April 12, 2023. We could use the first line of the CDX response to learn the first archived date and MIME type of the web page. For example, \texttt{www.example.com} was first archived on January 20, 2002 and is an HTML page (text/html MIME type). 

It is not feasible to work with the CDX index of the entire holdings of the Wayback Machine, which contains hundreds of billions of URLs, due to the immense scale of the data. To address this challenge, IA employs the ZipNum index as a scalable and efficient alternative for querying archived web resources. The ZipNum index, also known as the Ziplines cluster, provides a structured and compressed solution to accessing large CDX files \cite{aaron2013ZipNum, pywb_indexing, webarchive_indexing}. Unlike the traditional approach of performing a binary search across a single CDX file, which becomes computationally expensive and inefficient for archives at terabyte or petabyte scales, the ZipNum index organizes CDX data into gzipped clusters. Each cluster compresses approximately 6,000 CDX lines, which reflects the Internet Archive’s configuration at the time we received the IA ZipNum index sample file. This reduces the overall data footprint while preserving the ability to locate specific records quickly. Conceptually, the ZipNum index functions as a kind of skiplist over a much larger CDX index, enabling fast traversal by allowing lookups to skip over large portions of data (Figure~\ref{img:SamplingZipNum}). The IA's ZipNum index, though sampled from the CDX index, differs significantly in structure and purpose. Unlike CDX, which is typically used for full archival listings and replay services, ZipNum prioritizes efficient retrieval by using a subset of CDX data, structured for quick range queries~\cite{aaron2013ZipNum}. Each ZipNum entry contains a URL key (SURT), timestamp, CDX file part, byte offset, record size in bytes, and a block number, which collectively allow direct access to archived content within WARC/ARC files  (Figure~\ref{fig:ZipNum_snippet}). It is lexicographically sorted with the SURT as the primary key, the memento's datetime as the secondary key, and additional attributes following in order. This organization enables a high-level mapping of URLs, making retrieval efficient by ensuring only the necessary compressed clusters are accessed during a lookup. The ZipNum index thus significantly reduces computational overhead and storage requirements during data retrieval.

\begin{figure}[ht]
\begin{lstlisting}[numbers=none, backgroundcolor = \color{white}]
asia,city-sat)/thread28004.html 20130622195420	part-a-00001	3237741936	263882	999981
asia,citygolfcambodia)/site 20180811181040	part-a-00001	3239378785	332036	999987
asia,cityrental)/ 20130804002034	part-a-00001	3239987800	278249	999989
asia,citytours)/robots.txt 20160404154855	part-a-00001	3240266049	304687	999990
asia,civilengineer)/tag/kerinci 20101006182932	part-a-00001	3240570736	305894	999991
\end{lstlisting}
 \caption{A snippet of IA's ZipNum Index, showcasing its structured format with fields including SURT key, capture datetime, CDX file part, byte offset, record size, and block number. This format enables efficient lookup and retrieval of archived web resources.}
\label{fig:ZipNum_snippet}
\end{figure}

\section{Methodology}\label{sec:method}

\begin{figure}
  \centering
\frame{\includegraphics[width=\linewidth]{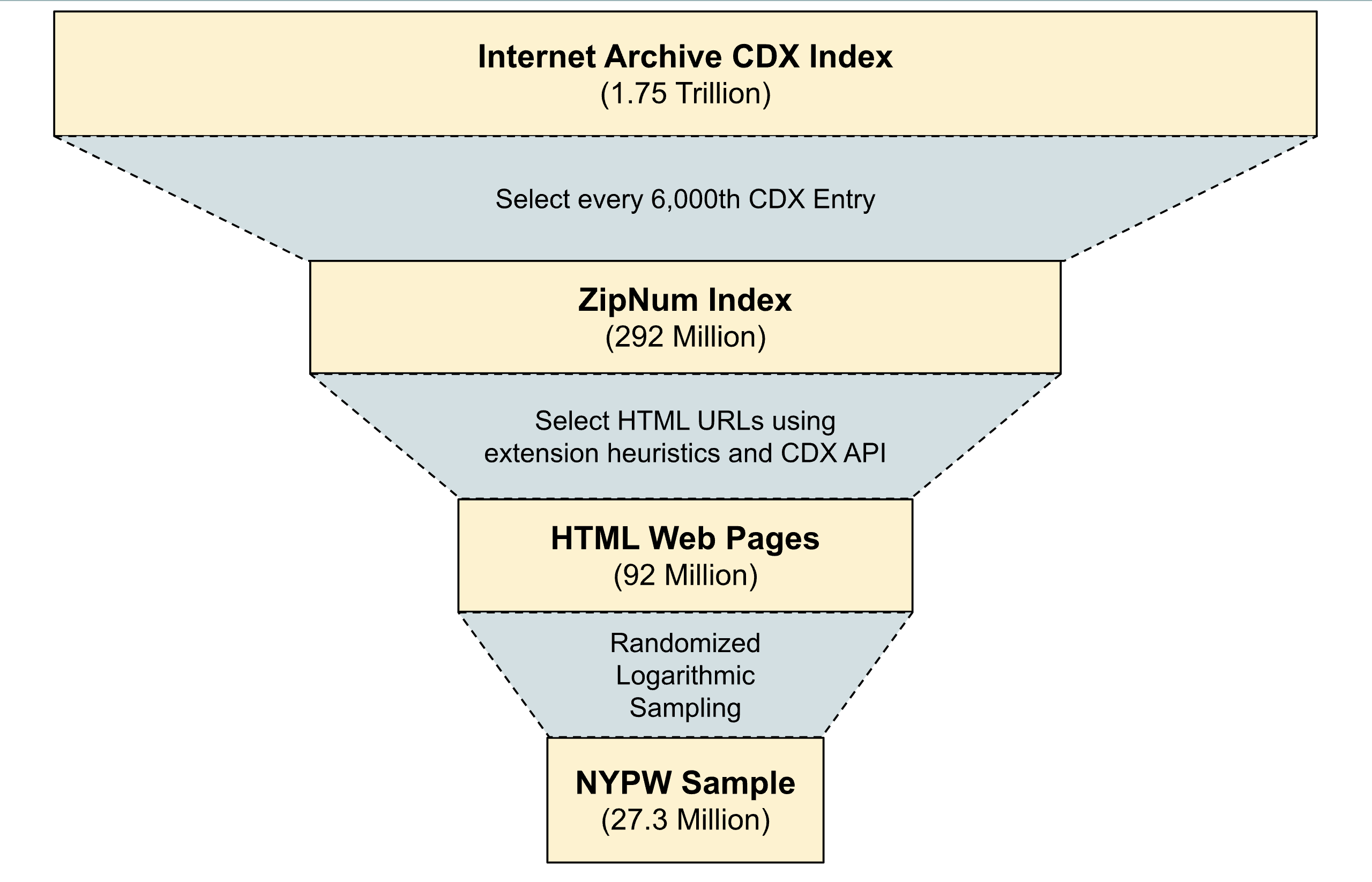}}
  \caption{The overview of NYPW sampling process illustrating the multi-step sampling methodology used to derive the NYPW Sample from the Internet Archive. The IA's CDX index contains 1.75 trillion entries from which every 6,000th CDX entry was selected to form ZipNum index, comprising 292 million entries. The next phase involved the selection of HTML URLs from ZipNum index using extension heuristics and the CDX API, narrowing the dataset to 92 million HTML web pages. Finally, through randomized logarithmic sampling, the process delivered the NYPW Sample, consisting of 27.3 million URLs (URI-Rs).
}
  \label{img:Samplingoverview}
\end{figure}

We acknowledge that creating a perfectly “representative sample of the web” is infeasible due to the web's vast diversity. To address this, we adopted a set of orthogonal dimensions to guide our sampling process. We considered time first archived as a proxy for when a web page was created. We selected URLs first archived across the 26 years of the IA collection, spanning 1996--2021. While URLs from the 1990s are sparse and not necessarily representative of the modern web, their inclusion captures the public’s attention and provides historical context.
We focused on \texttt{text/html} MIME type. Although a variety of MIME types exist in the Wayback Machine (e.g., PDFs, images, CSS), we prioritized HTML pages, as they align with public intuition of what constitutes “the web.” 
We sampled URLs from the IA’s ZipNum index file and employed various sampling techniques to ensure variety in domain and temporal representation. Our final NYPW sample consists of TimeMaps for 27.3 million URLs from 1996 to 2021, encompassing 7 million unique hosts. Figure~\ref{img:Samplingoverview} presents a funnel-shaped flowchart that illustrates the step-by-step filtering process from the full CDX index of IA to the final NYPW sample. The blue sections indicate the processes used for filtering and downsampling, while the yellow sections denote the resulting outputs at each stage. As described in section~\ref{sec:background}, the IA's ZipNum index is generated by the IA by selecting every 6,000th entry from the much larger CDX index of IA (1.75 trillion entries). After receiving the ZipNum file (containing 292 million entries), we first filtered it to isolate likely HTML URLs using extension-based heuristics. This step helped reduce the number of queries needed for the CDX API, as querying can be resource-intensive. Then, we used the CDX API to verify the MIME type of these URLs, resulting in a set of 92 million HTML web pages. To ensure variety and proportional representation across the dataset, we applied randomized logarithmic sampling to downsample the HTML URLs. From this, we produced the final NYPW sample, consisting of 27.3 million URLs. The remainder of this section will discuss each of these steps in detail.

\subsection{Sampling URLs from the ZipNum Index}\label{subsec:ZipNumsampling}
\begin{figure}
  \centering
  \frame{\includegraphics[width=\linewidth]{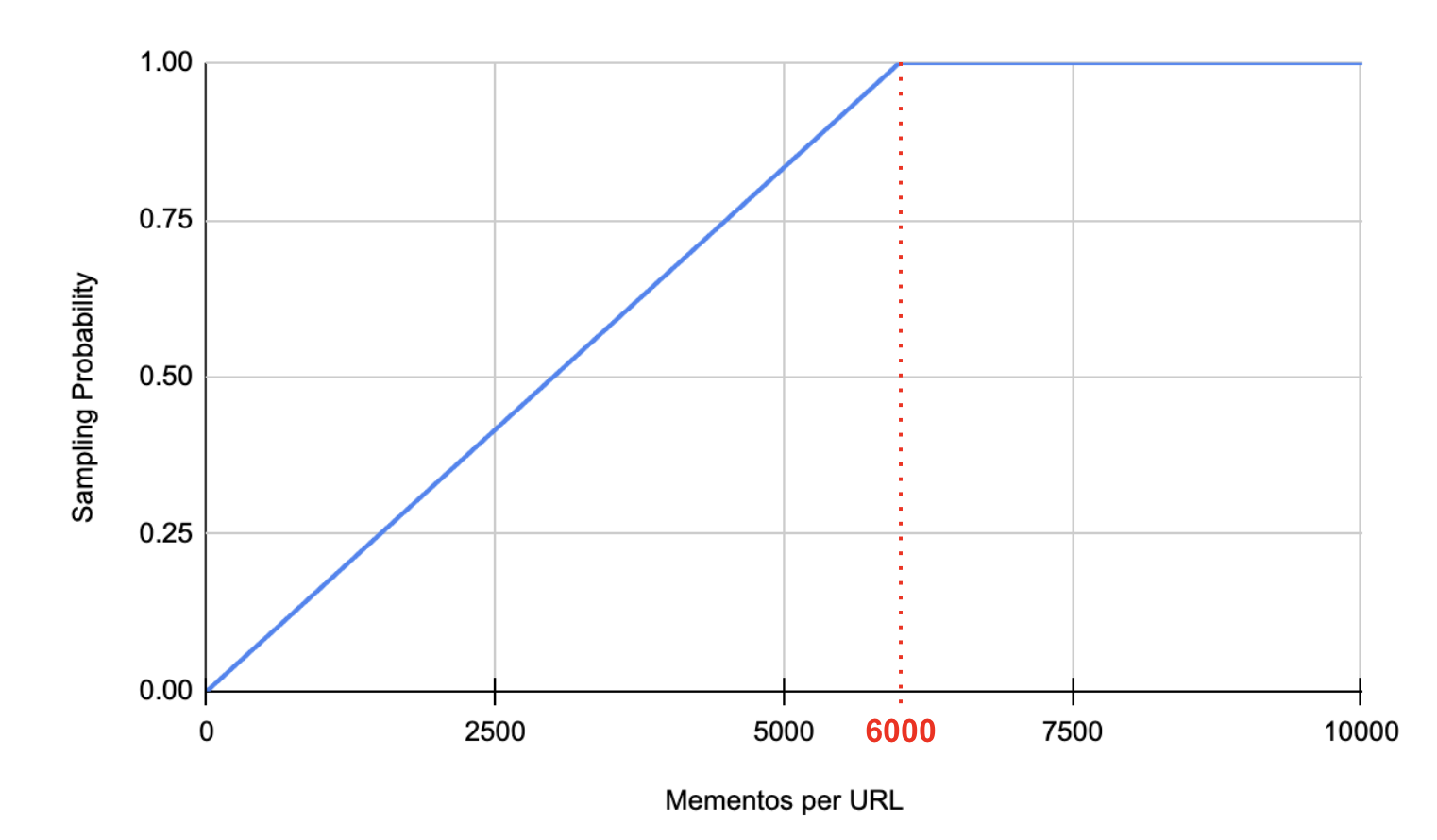}}
  \caption{The probability of a URL from the CDX index of the Wayback Machine getting sampled via the ZipNum index is linearly proportional to its memento count, with URLs containing more than 6,000 mementos being guaranteed inclusion in the ZipNum index.}
  \label{img:ZipNumsamplingprob}
\end{figure}

In August 2021, we received a ZipNum index file from the IA. The file contained 292 million records. After extracting the SURTs and removing duplicates, we were left with 285 million unique SURTs. Because the ZipNum file contains every 6000th line from the entire CDX index of IA, which is ordered by URL, sampling from it introduces a potential bias towards URLs with higher memento counts~\cite{kelly2017impact}. As illustrated in Figure~\ref{img:ZipNumsamplingprob}, the likelihood of a URL being sampled correlates with its number of mementos. Specifically, URLs with over 6000 mementos are certain to be included in the ZipNum index, resulting in a higher probability of being part of our sampled dataset. Consequently, URLs with a large number of mementos in the Wayback Machine (exceeding 6000) are more likely to be included in the sample, whereas URLs with fewer mementos have a reduced chance of selection. This bias at the URL level naturally extends to the domain level, given the hierarchical structure of the web. URLs under certain domains might be archived more frequently due to the overall popularity or perceived importance of the domain.

\begin{figure}
  \centering
\frame{\includegraphics[width=\linewidth]{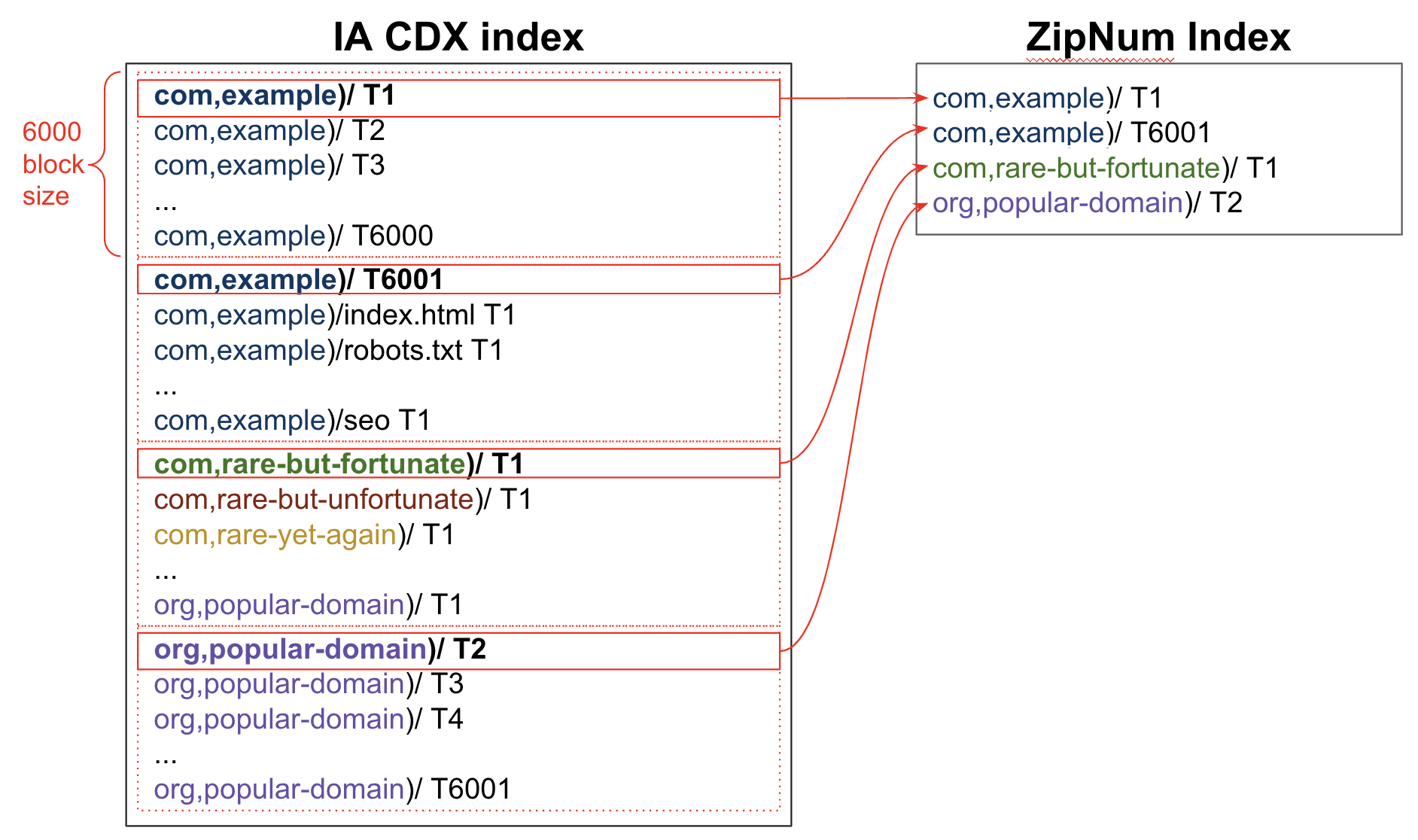}}
  \caption{Visualization of a ZipNum file creation from a CDX index. Popularly archived domains are more likely to be sampled in the ZipNum index}
  \label{img:SamplingZipNum}
\end{figure}

Figure~\ref{img:SamplingZipNum} shows the process of creating a ZipNum file from a CDX index, highlighting how popular domains with frequent archival are more likely to appear in the ZipNum index. The CDX index (left) stores URL entries and timestamps in blocks of 6000 lines, with each entry formatted as "URL $T_n$" (where $T_n$ represents a specific timestamp). Domains like \texttt{example.com} that are frequently archived have multiple timestamps associated with them, spanning several CDX blocks. The domain \texttt{example.com} is not just a commonly used placeholder but is explicitly reserved by RFC 2606~\cite{rfc2606} to provide a standardized example domain for documentation and illustrative purposes. Because they appear frequently, popular domains like these have a much higher likelihood of being sampled into the ZipNum index. In contrast, less popular domains, represented by hypothetical examples in Figure~\ref{img:SamplingZipNum} such as 
\texttt{com,rare-but-fortunate}; \texttt{com,rare-but-unfortunate}; and \texttt{com,rare-yet-again}, are archived less often and thus have a lower likelihood of being selected for the ZipNum index. Notably, \texttt{com,rare-yet-again} was sampled into the ZipNum index in this example, though it has only a single archived URL in the CDX. It was ``lucky'' to be sampled because it happened to be the top URL in a 6000-line CDX block, which allowed it to be represented despite its low frequency. Other low-frequency domains, like \texttt{com,rare-but-unfortunate} and \texttt{com,rare-yet-again}, are not in the ZipNum index due to their low archival frequency, making them less likely candidates for sampling.

\subsection{Filtering Invalid URLs}\label{subsec:invalidurls}
 We converted 285 million unique SURTs from the index file into standard URLs to facilitate analysis. The ZipNum index file contains a diverse array of URLs. Table~\ref{tab:allURLs} provides examples that illustrate the range of URLs encountered, including both valid and invalid cases. For example, URLs such as
 
 \texttt{https:///?dn=renunciationguide.com\&flrdr=yes\&nxte=css} and \texttt{https://*/robots.txt} in  Table~\ref{tab:allURLs} are invalid URLs. Specifically, the first URL lacks a domain name, while the second URL improperly uses a wildcard character (\texttt{*}) in the domain field. We filtered such malformed entries, which excluded 2.3 million likely invalid URLs, reducing the dataset to 282.7 million URLs deemed valid and usable for further analysis.
 
Upon closer inspection of the invalid URLs, we discovered that a significant portion of the issues stemmed from errors in the original SURT formatting within the ZipNum index file. Given the scale of the CDX index of IA and the potential for erroneous input, it is inevitable that some SURTs derived from IA's CDX index URLs also contain errors. Since the SURTs were incorrectly formatted, the conversion process resulted in invalid URLs. For instance, the file contained an invalid SURT \texttt{amazon.com/review/rs8o6bnbx9o5k}; the proper representation should be:
 \texttt{com,amazon)/review/rs8o6bnbx9o5k}.

\begin{table}[]
\begin{tabular}{|l|l|} \hline
\textbf{URLs}                                                                           & \textbf{Type}                       \\ \hline
https://brs53.dx.am/scripts/jquery.min.js                                               & Javascript                          \\ \hline
https://174.127.81.0/t/87/73/25/1-320x240.jpg                                           & Image                               \\ \hline
https://mf.ag/2121\_de.gif?exp=24559886473100                                           & Image                               \\ \hline
https://notiche.com.ar/index.php?limitstart=42                                          & HTML                                \\ \hline
{\color[HTML]{FE0000} \textit{https:///?dn=renunciationguide.com\&flrdr=yes\&nxte=css}} & {\color[HTML]{FE0000} \textit{N/A}} \\ \hline
{\color[HTML]{FE0000} \textit{https://*/robots.txt}}                                    & {\color[HTML]{FE0000} \textit{N/A}} \\ \hline
\end{tabular}
\caption{ZipNum index containing all kinds of URLs, including JavaScript, HTML, images, and other content types. Invalid URLs are highlighted in italics and red.}
\label{tab:allURLs}
\end{table}

\subsection{Filtering Likely URL Aliases}\label{subsec:urlaliases}
We targeted likely URL aliases~\cite{w3c_webarch,Bar-Yossef2009}, URLs that likely pointed to the same content but differed only in path parameters, such as session IDs. These session IDs are typically appended to URLs for tracking purposes but do not change the underlying content of the page. For example, on \texttt{clickbank.com}, the base URL \texttt{https://clickbank.com/index.html} appeared multiple times, each with a different session ID attached, such as \texttt{;jsessionid=09020c463c1db67320a9e4b9f65bf619}. While these session IDs result in distinct URLs, they do not affect the actual content of the page. The CDX server generally treats URLs with different session IDs as unique, resulting in separate TimeMaps for URLs that differ only by session parameters, such as \texttt{jsessionid}. However, the IA addresses this redundancy in their SURT library, specifically through the URLRegexTransformer class. This class includes regular expressions designed to detect and strip session identifiers from URLs, effectively canonicalizing them.\footnote{URL Regex Transformer: \url{https://github.com/internetarchive/surt/blob/master/surt/URLRegexTransformer.py}} 
To address this issue, we employed the same regex pattern on our dataset to identify the URLs with session IDs (Figure~\ref{fig:regex_sessionid}). After identifying the URLs with session IDs, we removed 12,996 such URLs from the dataset, ensuring that only unique URLs remained.

\begin{figure}[ht]
\begin{lstlisting}[numbers=none, backgroundcolor = \color{white}]
QUERY_SESSIONID = [
    re.compile("^(.*)(?:jsessionid=[0-9a-zA-Z]{32})(?:&(.*))?$", re.I),
    re.compile("^(.*)(?:phpsessid=[0-9a-zA-Z]{32})(?:&(.*))?$", re.I),
    re.compile("^(.*)(?:sid=[0-9a-zA-Z]{32})(?:&(.*))?$", re.I),
    re.compile("^(.*)(?:ASPSESSIONID[a-zA-Z]{8}=[a-zA-Z]{24})(?:&(.*))?$", re.I),
    re.compile("^(.*)(?:cfid=[^&]+&cftoken=[^&]+)(?:&(.*))?$", re.I),
    ]
\end{lstlisting}
 \caption{Regex pattern used to detect URLs with session IDs}
\label{fig:regex_sessionid}
\end{figure}

We also focused on another set of URL aliases, URLs that point to the same base directory but are distinguished by different index files, such as "index.html", "index.php", and similar variants. For instance, several URLs from the domain \texttt{abc.es}, including \texttt{https://abc.es/}, \texttt{https://abc.es/index.asp}, \texttt{https://abc.es/index.html}, and others, represent different URL versions of the same content. To identify these URLs, we checked if the final segment of the URL path contained an index file. More specifically, we used a regular expression to match the last path of the URL against the pattern \texttt{re.compile(r'index.[a-zA-Z]+\$')}. If the last part of the URL matched this pattern, it indicated that the URL was an index file. We then flagged and separated these URLs from the main dataset.

Through this process, we identified and set aside 155,014 likely URL aliases, ensuring they were not included in the broader analysis. We removed the likely URL aliases. This step was crucial for reducing redundancy and ensuring that different representations of the same content were properly handled.

\subsection{Canonicalization in TimeMaps}\label{subsec:canon_index}
The IA does not canonicalize index aliases of URLs, meaning that even if two URLs, such as \texttt{https://cnn.com/} and \texttt{https://cnn.com/index.html}, are aliases for the same page, they are treated as separate entities with distinct TimeMaps. While modern web practices have largely shifted toward automatically redirecting index pages to the root URL, this was not always the case. Many older websites explicitly served index pages as separate resources, making it possible for an index alias to have been archived before the root URL. We sought to understand whether canonicalizing the root homepage URL and its index alias would be beneficial, particularly if it would change the recorded first archive date. For example, how many URLs had their index alias archived before the root page? 

To investigate this, we collected TimeMaps for URL aliases of 54,136 root URLs that were first archived in the year 2000. For example, for a root URL like \texttt{https://foo.com}, we retrieved TimeMaps for its variations, such as \texttt{https://foo.com/index.htm} and \texttt{https://foo.com/index.html}. The objective was to determine how canonicalizing these variations might affect the timestamp of the first memento capture. Specifically, we extracted the first Memento-Datetime for the root URL and its \texttt{.htm} and \texttt{.html} versions. We then compared the datetimes of the index alias page with the root page to determine how often the index page had an earlier datetime, thereby indicating a potential change in the first memento capture if canonicalization were applied.  
 
The analysis revealed that only 218 out of the 54,136 \texttt{.htm} URLs had a Memento-Datetime earlier than their corresponding root URLs. Of these, 217 had their first memento captured in 1999, and one in 1997. Similarly, 325 \texttt{.html} URLs had an earlier Memento-Datetime than their root URLs, with all these \texttt{.html} URLs being first archived in 1999, one year before their respective root URLs. Overall, when considering Memento-Datetime at a yearly granularity, only 0.4\% of \texttt{.htm} URLs and 0.6\% of \texttt{.html} URLs showed an earlier datetime than their root URLs. This suggests that the impact of canonicalizing index aliases on the first memento capture is minimal.

Expanding this analysis to the entire dataset would require substantial time and resources, including a fivefold increase in CDX queries and efforts to resolve soft 404 errors, all for a likely minimal impact on the results. Therefore, we opted not to proceed with client-side canonicalization. Instead, we discarded the index aliases of the URLs and retained only the root URL. This topic could be further explored as a separate study in the future.

\subsection{Filtering URLs for the HTML Pages}\label{subsec:likely_html}
The URLs in the ZipNum index included embedded resources such as images, CSS files, and JavaScript (Table~\ref{tab:allURLs}), which we needed to exclude to limit our sample to HTML pages intended for humans. Identifying the MIME type of each URL was necessary, as it distinguishes between web pages and other resources. The MIME type field is not available in the ZipNum file, requiring us to retrieve this information separately using CDX API. However, querying the API for 282 million URLs would have been prohibitively time-consuming. To reduce this expense, we first filtered the URLs to those likely representing HTML pages and then queried the CDX API for their MIME types.

\subsubsection{Identifying Likely HTML Pages by File Extension}
We predicted the likelihood of a URL being an HTML page by examining its structure. We flagged the URLs as likely web pages based on URLs ending in a trailing slash or with the following extensions commonly associated with HTML content: \texttt{.do}, \texttt{.php}, \texttt{.aspx}, \texttt{.cgi}, \texttt{.pl}, \texttt{.asp}, \texttt{.jsp}, \texttt{.cfm}, \texttt{.html}, and \texttt{.htm}. This includes HTML files that are returned by the server as well as server-side executed scripts that will return HTML to the client.\footnote{Verify HTML URLs Script: \url{https://github.com/oduwsdl/nypw/blob/main/code/dataset_creation/verifyHTMLURLs.py}}

Table~\ref{tab:heuristic} displays the example URLs corresponding to each heuristic. The extension string \texttt{.php[0-9]} covers all the different versions of PHP files. For example, \texttt{\url{https://0001ktr.co.kr/bbs/bbs.php3?bbs\_mode=list\_form\&com=bbs\&db=freeboard\&page\_num=16}} has a \texttt{.php3} file extension used for files written in PHP 3. 

While analyzing the URLs, we observed various HTML file extensions captured by the \texttt{.[a-z]html} heuristic. These extensions denote specialized or historical forms of HTML files. For instance, the \texttt{.shtml} extension is used for server-side includes (SSI). The \texttt{.phtml}, \texttt{.jhtml}, \texttt{.dhtml}, and \texttt{.xhtml} extensions are associated with PHP, Java, dynamic, and extensible HTML files, respectively. The \texttt{.mhtml} extension stands for MIME HTML, utilized for web pages archived as a single file. Additionally, extensions such as \texttt{.ihtml}, \texttt{.nhtml}, \texttt{.zhtml}, \texttt{.chtml}, \texttt{.ghtml}, \texttt{.bhtml}, \texttt{.rhtml}, \texttt{.thtml}, \texttt{.vhtml}, \texttt{.ehtml}, \texttt{.fhtml}, \texttt{.ahtml}, \texttt{.whtml}, \texttt{.lhtml}, \texttt{.khtml}, \texttt{.hhtml}, \texttt{.ohtml}, \texttt{.yhtml}, \texttt{.uhtml}, and \texttt{.qhtml}, although less common, may be used for specific projects or by certain organizations to indicate different types of HTML content or processing needs.

\begin{table}[]
\begin{tabular}{|l|l|}
\hline
\textbf{Heuristic}    & \textbf{Example URL}                                                           \\ \hline
trailing slash/no ext & https://www.youtube.com\textbf{/}                                                       \\ \hline
.do                   & http://example.com/register\textbf{.do}                                                 \\ \hline
.php{[}0-9{]}         & https://notiche.com.ar/index\textbf{.php}                                               \\ \hline
.aspx                 & https://cigaroasis.asia/contact\textbf{.aspx}                                          \\ \hline
.cgi                  & https://0009.ir/cgi-sys/suspendedpage\textbf{.cgi}                                      \\ \hline
.pl                   & https://007thunderballpoker.com/11-5g-suited-poker-chip/pai-gow-poker-rules\textbf{.pl} \\ \hline
.asp                  & https://0000028.cnelc.com/productshop/newpro\textbf{.asp}                               \\ \hline
.jsp                  & https://006bai.net/404\textbf{.jsp}                                                     \\ \hline
.cfm                  & https://001ok.com/adventure\_nz\textbf{.cfm}?nft=1\&p=4\&t=4                            \\ \hline
.{[}a-z{]}html        & https://city-sat.asia/thread28004\textbf{.html}                                         \\ \hline
.htm                  & http://1st-international.com:80/profiles/16/PersonalBO893\textbf{.htm}                  \\ \hline
\end{tabular}
\caption{Example URLs corresponding to different heuristics for identifying HTML pages. Each heuristic targets specific URL patterns that are indicative of HTML content.}
\label{tab:heuristic}
\end{table}

We applied heuristics based on known file extensions (Table~\ref{tab:heuristic}) and filtered the dataset, reducing it to 222 million URLs likely to be HTML pages. This approach minimized the number of URLs requiring CDX API queries. 
Upon further analysis of the 222 million filtered URLs, we observed that a substantial portion of these URLs originated from a relatively small number of popular domains. For example, domains such as \texttt{twitter.com} accounted for a significant portion of the URLs. To reduce the number of queries to the CDX server further, we excluded 115 million URLs associated with popular domains to be processed separately (as will be discussed in Section~\ref{subsec:popular}). This additional filtering step reduced the dataset to 107 million likely HTML URLs. These remaining URLs were then used for CDX queries to confirm their MIME types. We then proceeded to query the CDX server for these 107 million URLs to confirm their MIME types and ensure that they were indeed HTML resources.

\subsubsection{Determining URLs with \texttt{text/html} MIME Type Using the CDX API}

We began by selecting 40 million URLs from the 107 million likely HTML pages to query the CDX API. This initial step allowed us to evaluate the computational and time cost of querying the CDX API at scale. Additionally, it enabled us to assess the effectiveness of our extension-based heuristics for predicting HTML pages.

Using this sample of 40 million likely HTML URLs, we queried the CDX API to collect the first CDX entry for each URL and extract its MIME type. Our analysis revealed that the heuristics correctly identified 88.12\% of the sampled URLs as HTML. Since we initially selected URLs that were already expected to be HTML, our evaluation focused on extracting actual HTML pages from this set. As a result, our dataset does not contain \textit{true negatives} (correctly identified non-HTML pages) or \textit{false negatives} (HTML pages misclassified as non-HTML pages). Instead, our results reflect how well the heuristic method performed in identifying HTML pages among those already suspected of being HTML. We observed both \textit{false positives} (cases where the file extension predicted HTML but the CDX response did not return a \texttt{text/html} MIME type) and \textit{true positives} where the prediction and the CDX response aligned. However, we eliminated the false positives (based on the reported MIME type), so our final sample only includes valid HTML entries, ensuring that non-HTML pages were excluded from further analysis. Table~\ref{tab:heuristicaccuracy} provides a detailed breakdown of the number of URLs accurately identified as HTML versus those incorrectly classified. For example, 98.3\% of URLs with the \texttt{.htm} extension were correctly identified as HTML pages.

\begin{table}[]
\begin{tabular}{|l|r|r|r|r|}
\hline
\textbf{Heuristic} &
  \multicolumn{1}{l|}{\textbf{\begin{tabular}[c]{@{}l@{}}URLs Predicted\\ as HTML\end{tabular}}} &
  \multicolumn{1}{l|}{\textbf{\begin{tabular}[c]{@{}l@{}}True\\ Positive\end{tabular}}} &
  \multicolumn{1}{l|}{\textbf{\begin{tabular}[c]{@{}l@{}}False\\ Positive\end{tabular}}} &
  \multicolumn{1}{l|}{\textbf{Precision}} \\ \hline
trailing slash/no ext & 23.3M  & 19.5M  & 3.8M   & 83.7\% \\ \hline
.do                   & 74.5K  & 63.4K  & 11.1K  & 85.1\% \\ \hline
.php{[}0-9{]}         & 4.7M   & 4.2M   & 533.4K & 88.7\% \\ \hline
.aspx                 & 888.9K & 800.3K & 88.6K  & 90.0\% \\ \hline
.cgi                  & 422.9K & 381.2K & 41.7K  & 90.1\% \\ \hline
.pl                   & 88.0K  & 80.8K  & 7.2K   & 91.8\% \\ \hline
.asp                  & 1.3M   & 1.2M   & 84.4K  & 93.7\% \\ \hline
.jsp                  & 219.8K & 205.9K & 13.9K  & 93.7\% \\ \hline
.cfm                  & 229.1K & 221.5K & 7.5K   & 96.7\% \\ \hline
.{[}a-z{]}html        & 7.5M   & 7.3M   & 164.2K & 97.8\% \\ \hline
.htm                  & 1.6M   & 1.6M   & 27.0K  & 98.3\% \\ \hline
\textbf{Total}                  & \textbf{40.3M}  & \textbf{35.6M}  & \textbf{4.8M}   & \textbf{88.2\%} \\ \hline
\end{tabular}
\caption{Performance of different heuristics for predicting HTML pages. The table shows the number of URLs predicted as HTML, correct and incorrect predictions, and the precision for each heuristic based on 40 million predicted HTML pages.}
\label{tab:heuristicaccuracy}
\end{table}

Subsequently, we extended the process to the remaining 67 million URLs, querying the CDX API for their first entry and extracting the MIME type and the first archived date. On average, each CDX API query took 0.03 seconds, resulting in approximately 29 days to gather results for all 107 million URLs. 
From the CDX query responses, we determined that 92 million out of 107 million URLs had the \texttt{text/html} MIME type, confirming their HTML status. While multiple MIME types can represent web pages, we restricted our selection to \texttt{text/html}, the most common and widely used format for human-readable HTML documents. Overall, 86.12\% of the 107 million URLs were correctly predicted as HTML using our heuristics. Table~\ref{tab:heuristic92M} details the number of URLs captured by each heuristic employed in this process.

\begin{table}[]
\begin{tabular}{|l|r|}
\hline
\textbf{Heuristic} & \multicolumn{1}{c|}{\textbf{URLs Correctly Predicted as \texttt{text/html}}} \\ \hline
trailing slash/no ext & 52.5M  \\ \hline
.do                   & 123.6K \\ \hline
.php{[}0-9{]}         & 9.8M   \\ \hline
.aspx                 & 2.1M   \\ \hline
.cgi                  & 929.9K \\ \hline
.pl                   & 192.1K \\ \hline
.asp                  & 3.0M   \\ \hline
.jsp                  & 691.0K \\ \hline
.cfm                  & 561.7K \\ \hline
.{[}a-z{]}html        & 18.7M  \\ \hline
.htm                  & 4.0M   \\ \hline
                      & 92.6M  \\ \hline
\end{tabular}
\caption{Count of URLs correctly predicted as \texttt{text/html} by each heuristic. We applied these heuristics to filter them down to 107 million likely HTML pages. Using the CDX API, we verified that 92.6 million URLs were \texttt{text/html}. The table shows the distribution of correct predictions across the different heuristics.
}
\label{tab:heuristic92M}
\end{table}

The reason we filtered our sample to only "likely HTML pages" before making any queries was to minimize the high computational cost associated with querying the CDX API on a large dataset. To achieve this, we implemented an initial filtering step that identified likely HTML pages by analyzing URLs with specific characteristics, such as trailing slashes or known filename extensions (e.g., \texttt{.html}, \texttt{.htm}, or the absence of an extension indicating directory pages). While this simple method provided an effective way to approximate HTML content, we acknowledge that more advanced methods could yield greater precision in detecting HTML pages. After performing this initial filtering, we queried the CDX API and refined our dataset to 92 million pages confirmed to be actual HTML content. This subset of 92 million pages then served as the foundation for our subsequent analyses and processing steps.

\subsection{Determining Year First Archived}\label{subsec:first_memento}
To study the lifespan of a webpage, it is crucial to determine its creation date~\cite{Hany2013Carbondate, TweetedAt:NaumanSawood}. Since the precise creation date of a webpage is often unavailable, we utilized the datetime of the earliest memento as an estimate for this purpose. The earliest memento serves as a reasonable proxy for the creation date, representing the first instance when the webpage was archived by the IA's Wayback Machine.

Our initial objective was to create a sample of 26 million URLs, with 1 million URLs first archived in each of the 26 years of the IA's existence, from 1996 to 2021. However, obtaining a chronological listing of all URLs based on their initial appearance is not straightforward using the IA’s Wayback Machine. This difficulty arises because the Wayback Machine’s index is organized by URLs as the primary key rather than by dates. Consequently, we could not directly retrieve a list of URLs sorted by their first archived date. To overcome this challenge, we queried the CDX API for the first record of the 92 million HTML URLs to identify the year they were first archived. To receive the first record from the CDX server, we could add the query parameter \texttt{\&limit=1} to our CDX query.\footnote{First CDX Response Script: \url{https://github.com/oduwsdl/nypw/blob/main/code/dataset_creation/firstCDXresponse.sh}} This process allowed us to accurately identify the earliest archived instance for each URL, enabling us to estimate their creation dates effectively and build a sample of the web for our study.

\begin{figure}
  \centering
  \frame{\includegraphics[width=\linewidth]{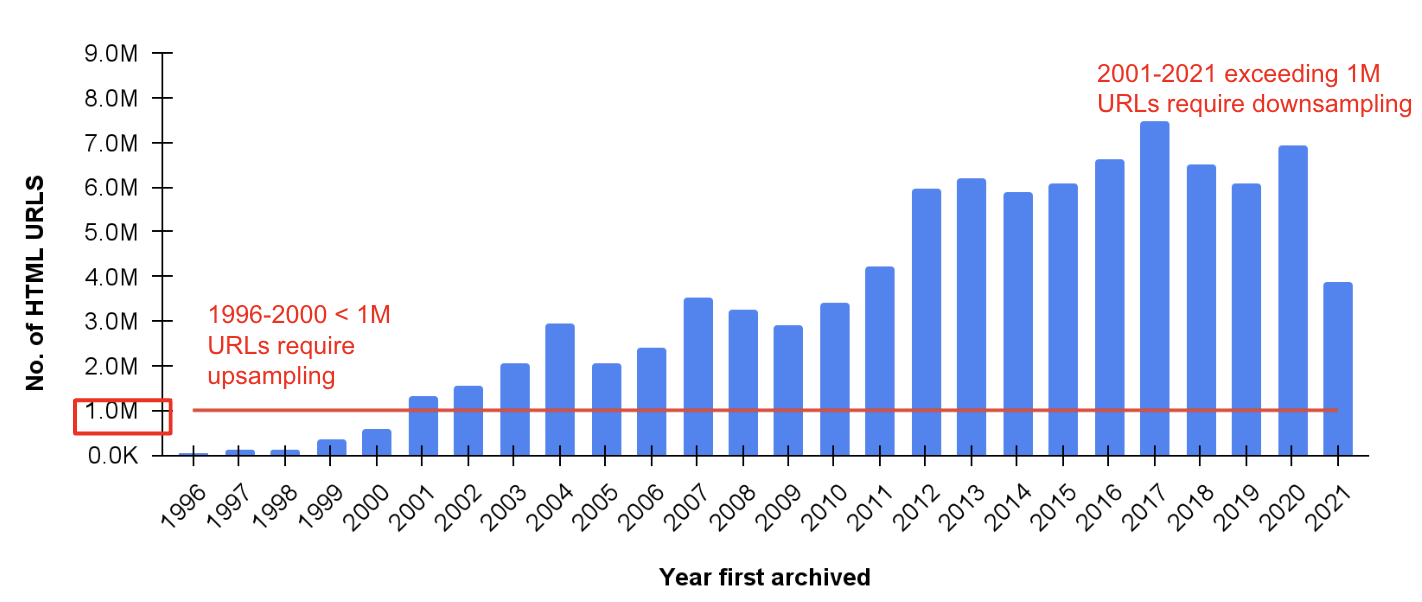}}
  \caption{The first archive year of 92 million sampled HTML URLs.}
  \label{img:firstarchived92m}
\end{figure}

We observed that a handful of URLs had their first archived date before 1996. For example, the earliest memento was for \texttt{https://cevi.be/}, with a Memento-Datetime of 1979-12-31. This was impossible,  so we removed URLs with the first archived dates before 1996. Figure~\ref{img:firstarchived92m} shows the distribution of the year of first capture for our 92 million URLs. We can see an increase in the number of archived URLs over the years depicts the significant increase in web and archiving capacity over time.  It is important to note that the year 2021 includes only eight months of data because the index file used is from August 2021. Similarly, 1996 has partial data since the IA's Wayback Machine began in May 1996. Consequently, both years show fewer URLs.

\subsection{Upsampling Early Years}\label{subsec:upsample}
As anticipated, and as shown in Figure~\ref{img:firstarchived92m}, we found fewer URLs first archived in the early years (1996--2000) compared to later years. To reach our target of 1 million URLs each year, we needed to identify more URLs that might have been first archived in these early years. We divided our 92 million URL sample into 2.2 million top-level URLs/root URLs and 89.9 million deep links. We observed that the early years contained more root URLs than later years. In contrast, we noticed an increase in the number of deep links archived over the years. We decided to extract the root URLs from the deep links to determine if this approach would help us upsample the earlier years. 

We identified the 20.4 million domains in the 92 million sample that had no root URLs in our sample. We extracted hostnames to form root URLs and then added these missing root URLs to our sample. 
For example, we extracted \texttt{https://reddit.com/} from \texttt{\url{https://reddit.com/r/argentina/comments/1ruebz/cient\%c3\%adficos\_chubutensesi}}. 

Next, we queried the IA's CDX API for the newly identified 20.4 million root URLs. After filtering out the 332,000 root URLs that received empty CDX responses (indicating that they were not archived), we were left with 20.1 million new root URLs. We then retrieved the earliest Memento-Datetime for these newly added URLs.

\begin{figure}
  \centering
  \frame{\includegraphics[width=\linewidth]{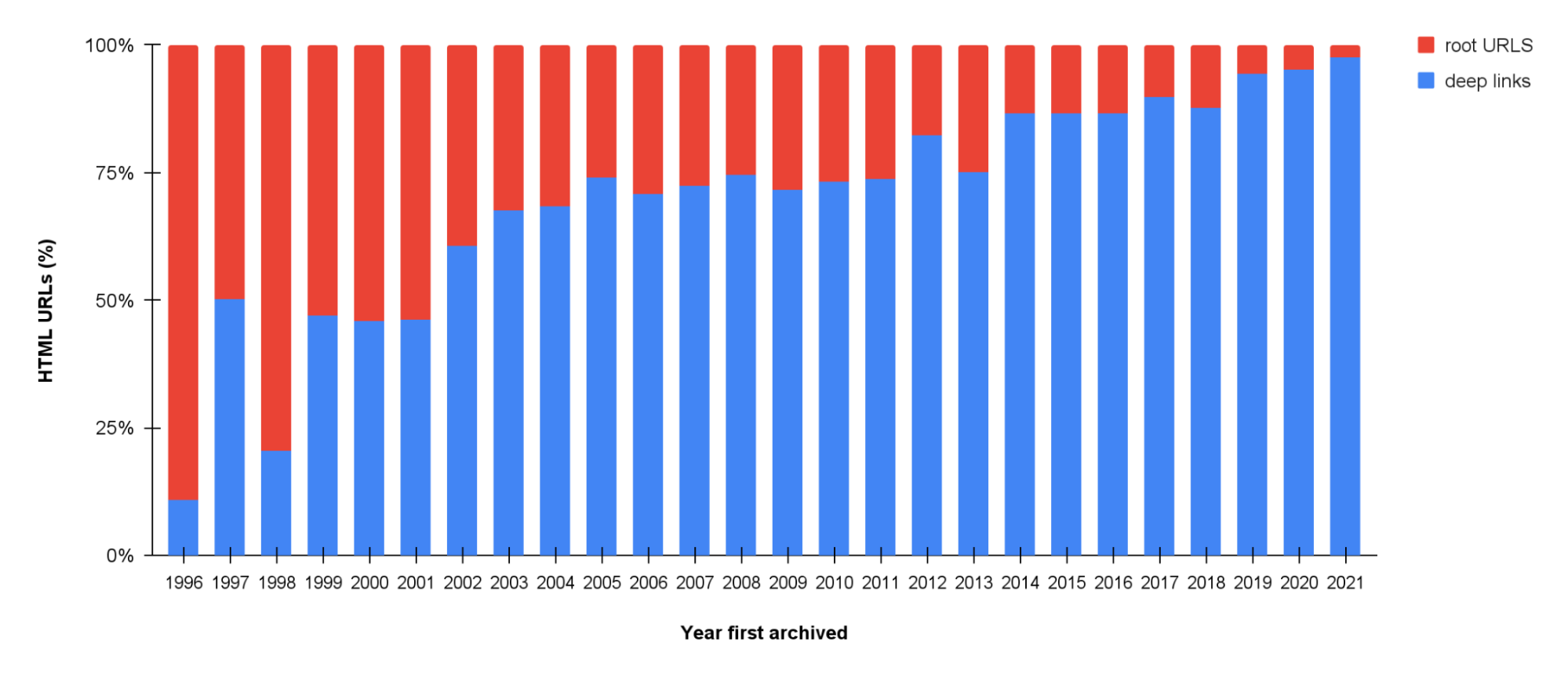}}
  \caption{The first archive year of 22.3M root URLs and the 89.9 million deep links. }
  \label{img:firstarchivedupsample}
\end{figure}

By combining these 20.1 million new root URLs with the 2.2 million originally present in our 92 million sample, we received a total of 22.3 million root URLs. Figure~\ref{img:firstarchivedupsample} illustrates the distribution of root URLs across the early years. We decided to retain the root URLs that were first archived between 1996 and 2002. The number of URLs approximately doubled for the 1996--2002 samples after adding new root URLs in the dataset. 

Upsampling enabled us to increase the number of URLs from the early years, which is of greater interest to our study. However, even with this upsampling of root URLs, we did not achieve our target of 1 million URLs per year in the early years. Consequently, we adjusted our goal and clustered the early years (1996--2000) to reach a total of 1.2 million URLs collectively.

\subsection{Handling Popular Domains}\label{subsec:popular}
Popular domains tend to be archived more frequently due to their high visibility and the large number of inbound links directing traffic to them. These domains, often associated with well-known websites or high-traffic platforms, appear more frequently in hyperlink networks, increasing the likelihood that web crawlers will discover and revisit them. Their dense interconnections within the web’s link structure make them more likely to be encountered by crawlers. Inbound links, or backlinks, are hyperlinks from other websites that point to a particular page, effectively increasing its discoverability. This process reflects the principles of PageRank~\cite{brin1998anatomy}, where pages with more inbound links are considered more important and are more likely to be traversed by crawlers. As a result, popular domains benefit from higher archival coverage, while less-linked pages may be overlooked. Additionally, users are more likely to submit URLs from popular domains to web archiving services, such as Save Page Now~\cite{Graham2019}, either to preserve important content or because these sites are more widely recognized and shared. As a result, they generate a larger number of mementos, capturing a more extensive historical record of their content. This frequent archiving ensures that significant changes, updates, and events related to these popular domains are preserved, providing a rich resource for future research and analysis. However, relying on a sample where most data is concentrated on just a few popular domains does not provide a balanced or comprehensive view of the web. In this section, we outline the methods we employed to address this issue in the archived index data, ensuring a more diverse sample by mitigating the over-representation of these heavily archived domains. 

To minimize the number of queries made to the CDX server, we initially excluded 111 million URLs associated with popular domains like \texttt{twitter.com}. Our priority was to collect the first CDX entry for domains that were not overly represented in the dataset. Once this data was collected, we reintroduced the previously excluded domain into our sample. For example, we employed a randomized sampling process for \texttt{twitter.com} URLs to ensure a balanced representation across all years, collecting at least 20 URLs for each year from 2016 to 2021. This threshold aligns with the approximate cap for popular sites during these later years. We used a file of known \texttt{twitter.com} URLs as our input and randomly selected one URL at a time for analysis. For each selected URL, we queried the Internet Archive’s CDX API to obtain the timestamp of its earliest archived capture and extracted the year from this timestamp. We set up separate counters for each year from 2016 to 2021 and incremented the relevant counter whenever a URL’s first archived date matched one of these years. We repeated this sampling process until each yearly counter reached at least 20 URLs.\footnote{First Archive Popular Domain Script: \url{https://github.com/oduwsdl/nypw/blob/main/code/dataset_creation/first_archive_twitter.sh}} 

To estimate the ranking of \texttt{twitter.com} and other prominent domains in the archive, we compared the number of Twitter URLs with those from other top domains for the years 2016--2021. Based on Twitter's rank within the top domain list for each year, we strategically downsampled the number of Twitter URLs to maintain a consistent and proportional representation in the final dataset. To downsample, we applied the logarithmic-scale downsampling technique, as detailed in Section~\ref{subsec:balance_domains}, to ensure a balanced sample.

\subsection{Balancing URLs per Domain}\label{subsec:balance_domains}

We observed that in each yearly sample, nearly 70\% of the domains contained only a single URL, revealing a long tail of low-coverage domains. For example, in the 2016 sample, 1.7 million domains (79\%) had just one archived URL. The long tail of URLs features domains that are likely not part of most users' experience, though we cannot be certain for foreign sites. For instance, domains such as \texttt{000246.com}, \texttt{0000k.com}, \texttt{00000000000.cn}, or \texttt{zzz7.com} seem unlikely to be widely recognized or frequently visited, whereas foreign domains like \texttt{blumen-konzelmann.de} may hold relevance within specific linguistic or cultural contexts. We chose to reduce these single-URL domains in our sample because their limited archival presence suggests that they are either rarely visited, short-lived, or not part of the mainstream web experience. Additionally, many of these domains may be artifacts of large-scale crawling rather than reflective of meaningful user activity. As detailed below, we examined the number of unique domains in each yearly sample and reduced the long tail by 90\%. 

We observed that certain popular domains, such as \texttt{google.com} and \texttt{github.com}, were over-represented in our sample. This over-representation occurs because popular domains tend to be archived more frequently. As shown in Table~\ref{tab:Top20beforedownsampling}, the top 20 domains with the highest number of URLs in our 92 million URLs sample illustrate this trend. To address this imbalance, we developed a logarithmic-scale downsampling method, represented by Equation~\ref{eq:downsampling}, to reduce the over-sampled domains:

\begin{equation}
\text{Reduced URLs} = \min \left( N, \; \mathrm{round}\left( K \cdot \log(N) + C \right) \right)
\label{eq:downsampling}
\end{equation}

\begin{itemize}
    \item \(N\) is the total number of URLs from a given domain.
    \item \(K\) is a constant that controls the rate of reduction.
    \item \(C\) is the minimum number of URLs that should always be retained.
\end{itemize}

The term \(\log(N) + C\) sets a baseline for URL selection based on a logarithmic scale. Since the logarithm grows slowly as \(N\) increases, this means that for domains with many URLs, the number of retained URLs increases gradually rather than linearly. The addition of \(C\) ensures that even domains with very few URLs retain at least \(C\) samples. The full term \(K \cdot \log(N) + C\) further scales this selection by a factor of \(K\), providing more control over how aggressively URLs are downsampled. The \(\min\) function ensures that if this calculated number exceeds \(N\), all URLs are kept. Equation~\ref{eq:downsampling} ensures that highly populated domains contribute fewer URLs to the dataset while still preserving a meaningful sample, preventing over-representation of certain domains while maintaining diversity in the dataset.

For example, in the case of \texttt{github.com}, which had 1.3 million URLs in our dataset, the sheer volume of URLs offered diminishing returns. Adequate coverage of \texttt{github.com} could be achieved with a significantly smaller subset of URLs. We applied Equation~\ref{eq:downsampling} to each domain in the yearly samples.
Our goal was to reduce the number of URLs while maintaining approximately 1 million URLs per yearly sample. To achieve this, we adjusted the parameters $K$ and $C$ in Equation~\ref{eq:downsampling}, allowing us to fine-tune the downsampling process for each year's dataset.

\begin{table}[]
\begin{tabular}{|r|l|r|}
\hline
\multicolumn{1}{|l|}{\textbf{No.}} & \textbf{Domain} & \multicolumn{1}{l|}{\textbf{URLs}} \\ \hline
1  & google.com          & 1.8M   \\ \hline
2  & github.com          & 1.3M   \\ \hline
3  & reddit.com          & 1.1M   \\ \hline
4  & youtube.com         & 866.6K \\ \hline
5  & tumblr.com          & 685.3K \\ \hline
6  & wordpress.com       & 577.4K \\ \hline
7  & blogspot.com        & 521.2K \\ \hline
8  & yahoo.com           & 456.9K \\ \hline
9  & facebook.com        & 308.3K \\ \hline
10 & instagram.com       & 245.3K \\ \hline
11 & bebo.com            & 229.7K \\ \hline
12 & amazon.com    & 200.5K \\ \hline
13 & url.cn        & 196.2K \\ \hline
14 & webshots.com  & 191.6K \\ \hline
15 & twitpic.com   & 160.5K \\ \hline
16 & wikipedia.org & 147.7K \\ \hline
17 & webs.com      & 140.6K \\ \hline
18 & verizon.net   & 140.1K \\ \hline
19 & qq.com       & 139.0K \\ \hline
20 & hyves.nl     & 124.6K \\ \hline
\end{tabular}
\caption{Top 20 domains from the dataset of 92 million HTML URLs before downsampling }
\label{tab:Top20beforedownsampling}
\end{table}

An essential aspect of this process was balancing the reduction of URLs from the "long tail" — domains with only a single URL — against limiting the over-representation of popular domains. This balance is crucial for achieving an equitable distribution of URLs across all domains within a yearly sample. If we were to remove too much of the long tail, the dataset would contain fewer unique domains, and as a result, logarithmic downsampling would allocate a disproportionately higher number of URLs to the more popular domains. This would not only distort the sample within a given year but also lead to significant over-representation of these domains when comparing data across multiple years. On the other hand, if too much of the long tail were retained, many domains would have only minimal representation.

To address these concerns and ensure an even distribution of URLs across domains, we opted to reduce the long tail for years in which the number of unique domains exceeded 900,000. Specifically, we reduced the long tail by 90\% for the years 2004 and 2006–2020. The threshold of 900,000 domains was determined through a trial-and-error process, where we observed that retaining the full long tail for years with more than 900,000 domains skewed the distribution heavily towards the popular domains. For years where the number of domains fell below this threshold, we retained the long tail to avoid inflating the number of URLs allocated to the most popular domains. After adjusting the long tail, we applied logarithmic downsampling to achieve the target of approximately 1 million URLs per year while ensuring a more balanced representation across domains. This approach allowed us to mitigate the trade-offs between reducing the long tail and avoiding over-representation of popular domains, thereby maintaining a more accurate reflection of domain distribution across the yearly samples.

In our implementation of the logarithmic-scale downsampling, we set $C=1$, while $K$ was adjusted for each yearly sample. Table~\ref{tab:KandCvalues} provides the specific $C$ and $K$ values used for each year, along with the final number of URLs selected per year.

\begin{table}[]
\begin{tabular}{|l|l|l|l|l|}
\hline
\textbf{Year} & \textbf{No. of Domains} & \textbf{No. of URLs} & {\color[HTML]{FE0000} \textbf{K}} & {\color[HTML]{FE0000} \textbf{C=1, Reduced URLs}} \\ \hline
1996-2000 & 1.4M & 2.6M & {\color[HTML]{FE0000} \textbf{1}} & {\color[HTML]{FE0000} 1.7M} \\ \hline
2001 & 1.5M & 2.6M & {\color[HTML]{FE0000} \textbf{1}} & {\color[HTML]{FE0000} 1.8M} \\ \hline
2002 & 1.2M & 2.4M & {\color[HTML]{FE0000} \textbf{1}} & {\color[HTML]{FE0000} 1.6M} \\ \hline
2003 & 820.3K & 2.1M & {\color[HTML]{FE0000} \textbf{2}} & {\color[HTML]{FE0000} 1.3M} \\ \hline
2004 & 290.4K & 2.3M & {\color[HTML]{FE0000} \textbf{5}} & {\color[HTML]{FE0000} 1.2M} \\ \hline
2005 & 635.8K & 2.1M & {\color[HTML]{FE0000} \textbf{3}} & {\color[HTML]{FE0000} 1.2M} \\ \hline
2006 & 319.1K & 1.8M & {\color[HTML]{FE0000} \textbf{5}} & {\color[HTML]{FE0000} 1.2M} \\ \hline
2007 & 369.4K & 2.8M & {\color[HTML]{FE0000} \textbf{2}} & {\color[HTML]{FE0000} 1.1M} \\ \hline
2008 & 318.1K & 2.6M & {\color[HTML]{FE0000} \textbf{3}} & {\color[HTML]{FE0000} 1.2M} \\ \hline
2009 & 287.8K & 2.3M & {\color[HTML]{FE0000} \textbf{5}} & {\color[HTML]{FE0000} 1.2M} \\ \hline
2010 & 403.5K & 2.6M & {\color[HTML]{FE0000} \textbf{2}} & {\color[HTML]{FE0000} 1.2M} \\ \hline
2011 & 418.6K & 3.4M & {\color[HTML]{FE0000} \textbf{2}} & {\color[HTML]{FE0000} 1.3M} \\ \hline
2012 & 483.7K & 5.1M & {\color[HTML]{FE0000} \textbf{1}} & {\color[HTML]{FE0000} 1.1M} \\ \hline
2013 & 463.0K & 5.2M & {\color[HTML]{FE0000} \textbf{1}} & {\color[HTML]{FE0000} 1.0M} \\ \hline
2014 & 370.6K & 5.1M & {\color[HTML]{FE0000} \textbf{2}} & {\color[HTML]{FE0000} 1.2M} \\ \hline
2015 & 421.0K & 5.1M & {\color[HTML]{FE0000} \textbf{2}} & {\color[HTML]{FE0000} 1.3M} \\ \hline
2016 & 605.4K & 5.1M & {\color[HTML]{FE0000} \textbf{1}} & {\color[HTML]{FE0000} 1.2M} \\ \hline
2017 & 531.1K & 6.4M & {\color[HTML]{FE0000} \textbf{1}} & {\color[HTML]{FE0000} 1.2M} \\ \hline
2018 & 440.2K & 5.6M & {\color[HTML]{FE0000} \textbf{2}} & {\color[HTML]{FE0000} 1.4M} \\ \hline
2019 & 314.9K & 5.4M & {\color[HTML]{FE0000} \textbf{3}} & {\color[HTML]{FE0000} 1.2M} \\ \hline
2020 & 382.7K & 5.8M & {\color[HTML]{FE0000} \textbf{3}} & {\color[HTML]{FE0000} 1.2M} \\ \hline
2021 & 501.9K & 3.9M & {\color[HTML]{FE0000} \textbf{2}} & {\color[HTML]{FE0000} 813.7K} \\ \hline
 &  &  & {\color[HTML]{FE0000} \textbf{}} & {\color[HTML]{FE0000} \textbf{27.3M}} \\ \hline
\end{tabular}
\caption{Yearly values of constants C and K, and the corresponding number of URLs selected per year after applying the downsampling equation.}
\label{tab:KandCvalues}
\end{table}

After applying downsampling, we obtained a file containing hostnames along with the corresponding count of unique HTML URLs, adjusted using logarithmic downsampling. To select URLs for each domain, we could either randomize the list and select a random subset of k URLs \footnote{Select Downsampled URLs: \url{https://github.com/oduwsdl/nypw/blob/main/code/dataset_creation/select_downsampled_domain_urls.py}} or apply reservoir sampling\cite{vitter1985random, alam_streamsampler}. For each domain, we first ensured the inclusion of the root URL and then randomly selected the remaining URLs based on the allocated count for that domain.

Downsampling the URLs flattened out the large discrepancy in the number of URLs per domain. Table~\ref{tab:Top20afterdownsampling} shows the ranking of domains based on the number of URLs after downsampling. The domain rankings before and after downsampling are highly correlated ($r = 0.994$, Pearson correlation coefficient). In other words, even though our equation does not strictly maintain the ordering after reducing the data, the rankings of the domains are nearly identical, with only minimal differences. For instance, \texttt{yahoo.com}, which was in 8th place before downsampling, is now ranked 1. After downsampling, we had 7 million unique domains in our sample. Ultimately, we re-adjusted our goal to a sample of 27.3 million URLs.
\begin{table}[]
\begin{tabular}{|r|l|r|r|}
\hline
\multicolumn{1}{|l|}{\textbf{No.}} & \textbf{Domain} & \multicolumn{1}{l|}{\textbf{URLs}} & \multicolumn{1}{l|}{\textbf{Reduced URLs}} \\ \hline
1  & yahoo.com     & 456.9K & 489 \\ \hline
2  & blogspot.com  & 521.2K & 463 \\ \hline
3  & google.com    & 1.8M   & 459 \\ \hline
4  & amazon.com    & 200.5K & 453 \\ \hline
5  & wikipedia.org & 147.7K & 439 \\ \hline
6  & house.gov     & 77.7K  & 417 \\ \hline
7  & msn.com       & 44.7K  & 401 \\ \hline
8  & yahoo.co.jp   & 49.2K  & 396 \\ \hline
9  & wordpress.com & 577.4K & 394 \\ \hline
10 & cnn.com       & 60.9K  & 391 \\ \hline
11 & ca.gov        & 40.0K  & 387 \\ \hline
12 & ebay.com      & 42.6K  & 383 \\ \hline
13 & go.com        & 39.1K  & 382 \\ \hline
14 & amazon.de     & 51.9K  & 382 \\ \hline
15 & microsoft.com & 34.4K  & 380 \\ \hline
16 & senate.gov    & 38.0K  & 379 \\ \hline
17 & sina.com.cn   & 33.5K  & 378 \\ \hline
18 & amazon.co.uk  & 37.3K  & 377 \\ \hline
19 & nih.gov       & 32.6K  & 376 \\ \hline
20 & amazon.co.jp  & 44.2K  & 376 \\ \hline
\end{tabular}
\caption{The top 20 domains from the dataset of 27.3 million HTML URLs after downsampling.}
\label{tab:Top20afterdownsampling}
\end{table}

\subsection{Collecting TimeMaps}\label{subsec:collect_timemaps}
We issued a CDX Pagination API query for the downsampled 27.3 million URLs to obtain their entire TimeMaps. The Pagination API  allows for sequential querying of CDX data, which helps gather metadata for URLs with large TimeMaps. Initially, we used the parameter \texttt{showNumPages=true} when querying the CDX API to determine the total number of pages for a URL. We then retrieved records from each page using the \texttt{\&page} parameter. After gathering the data, we merged these pages into a single file to create a TimeMap for a URL.\footnote{CDX TimeMaps Page Script: \url{https://github.com/oduwsdl/nypw/blob/main/code/dataset_creation/cdxtimemaps_page.sh}}

We also saved the HTTP response for each query made to the CDX API, which allowed us to recognize if a URL received any transient error and query for that again. We also analyzed the cost of collecting the TimeMap, which was approximately 0.07s/TimeMap. It took around 22 days to collect TimeMaps for 27.3 million URLs. These TimeMaps were 1.4TB in size.

\subsection{Revisit Records in TimeMaps}\label{subsec:revisit_record}
\begin{figure}
  \centering
  \frame{\includegraphics[width=\linewidth]{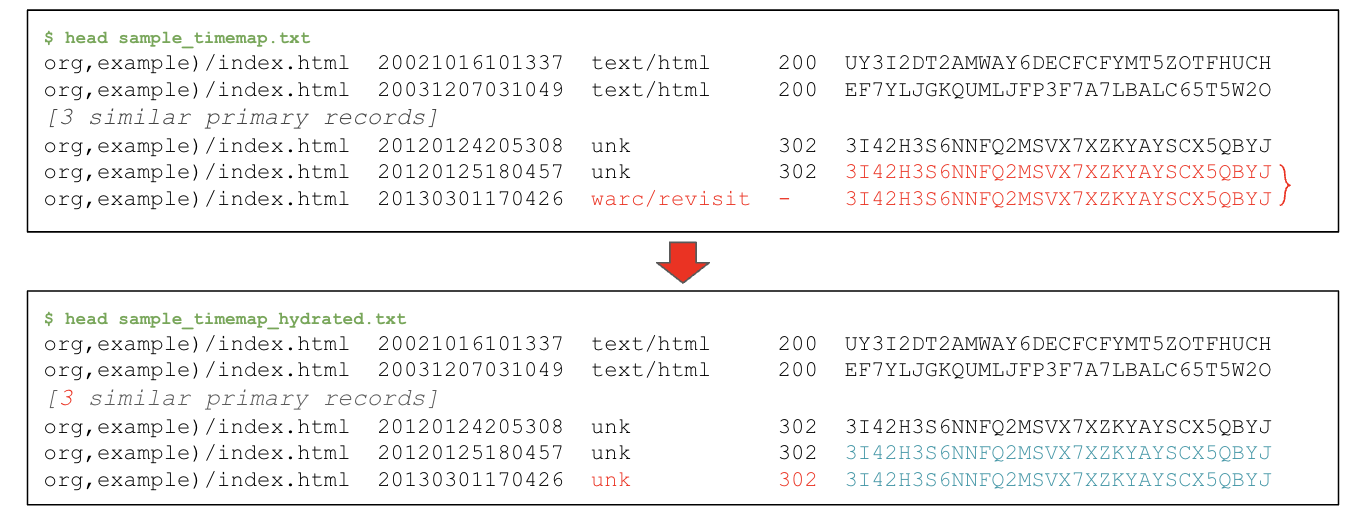}}
  \caption{Hydrating the `revisit' record in the TimeMaps. In the upper TimeMap snippet, \texttt{sample\_timemap.txt} displays entries for example.com, marked as "warc/revisit" for MIME type and lacking status codes marked by "\textbf{-}", indicating abbreviated content reliant on prior archived records. The lower snippet, \texttt{sample\_timemap\_hydrated.txt}, embodies the rehydration process where these missing status codes are restored by leveraging previous records with matching content digest. 
  }
  \label{img:hydratewarc}
\end{figure}

We noticed ``\texttt{warc/revisit}'' in the MIME type column and a dash (``\texttt{-}'') in the status code column for some rows of some of our collected TimeMaps. These records in the TimeMap are termed as a "revisit" record, which documents the revisitation of already archived content. It no payloads to reduce storage cost and these are created when an earlier related archival record has the exact same content\cite{warc_specification_1_1}. The status code of revisit records must be rehydrated from prior records for accurate lifespan analysis. The rehydration process involves utilizing the payload digest field of the CDX index to identify matching records from previous crawls. By comparing the digest of the current "warc/revisit" record with the digests of past archived records, we can locate the previous content that corresponds to the revisited entry. This ensures that the missing status codes, often omitted in revisit records, can be restored accurately. For example, in Figure~\ref{img:hydratewarc}, in the upper snippet of the TimeMap file, entries for "example.com" are marked as "warc/revisit" with MIME type information but lack associated status codes, indicated by the "-" symbol. This absence highlights that the content is abbreviated and dependent on prior archived records. In contrast, the lower snippet illustrates the rehydration process, where the missing status codes are restored by referring to the previous records with matching content digests. This allows for a complete and accurate representation of the web page's history, including the vital status codes that are necessary for lifecycle analysis.

We implemented a Least Recently Used (LRU) cache \cite{wikipedia_cache_replacement} to efficiently hydrate the ``\texttt{warc/revisit}'' records in our dataset.\footnote{Hydrate WARC Script: \url{https://github.com/oduwsdl/nypw/blob/main/code/dataset_creation/hydratewarc.py}} We analyzed a sample of 100 randomly selected TimeMaps to determine an optimal size for the LRU cache. Our primary goal was to ensure the cache could hold enough records to efficiently "rehydrate" WARC revisit records. To achieve this, we measured the number of records, or lines, between each WARC revisit record and its corresponding prior record within each TimeMap (Figure~\ref{img:hydratewarc}). Our analysis revealed a maximum distance of 792 lines between a revisit record and its associated prior record. Based on this data, we set the LRU cache's maximum size to 1,000 records, providing a buffer that accommodates this maximum distance and minimizes the risk of prematurely discarding necessary data.

The LRU cache is designed to retain only the most recently accessed records, automatically discarding the least recently used ones when it reaches capacity. This approach ensures that frequently accessed records remain available for immediate retrieval, significantly enhancing the efficiency of the rehydration process by reducing redundant lookups and minimizing retrieval times.

\subsection{Issues in TimeMaps}\label{subsec:issuesintimemap}

We analyzed the most frequently archived URLs in each yearly dataset and observed that some URLs with high memento counts ended with an asterisk (\texttt{*}). The asterisk is not a valid character in a URL and should be percent-encoded as \texttt{\%2A}, as specified in RFC~3986~\cite{rfc3986}. Specifically, we identified 2,757 such invalid URLs. Table~\ref{tab:asterisk} presents the top 5 of these 2,757 invalid \texttt{*} URLs with the highest reported memento counts. The high memento counts for these invalid URLs were due to the CDX server interpreting the asterisk as a wildcard. In CDX API queries, an asterisk at the end of a URL path is often used to match any URL that begins with the preceding string. For example, \texttt{https://mediafire.com/?8pzmr03tf9o*} would return TimeMaps for all URLs that start with \texttt{https://mediafire.com/?8pzmr03tf9o}, resulting in a TimeMap containing 135.9 million mementos.

To address this, we removed the TimeMaps associated with these invalid URLs ending with an asterisk. We then trimmed these invalid URLs to their hostnames, similar to the technique described in Section~\ref{subsec:upsample}. We checked these hostnames against our dataset and identified 672 new domains, for which we subsequently collected TimeMaps.

\begin{table}[]
\begin{tabular}{|l|r|}
\hline
                               & \multicolumn{1}{l|}{\textbf{Reported}} \\ 
\textbf{URL}  & \multicolumn{1}{l|}{\textbf{mementos}} \\ \hline
https://mediafire.com/?8pzmr03tf9o*          & 140.4M                                        \\ \hline
https://irishtimes.com/gadgets/..*           & 135.9M                                        \\ \hline
\begin{tabular}[c]{@{}l@{}}https://bdbiosciences.com/eu/solrsearch?page=1\&pgsize=50\&pgsize=\\ 50\&q=*:relevance:size:25+\%c2\%b5g:size:25+tests\&selectedtab=prod\\ uctcatalog\&sort=\&text=*\end{tabular} &
  118.2M \\ \hline
\begin{tabular}[c]{@{}l@{}}https://madmimi.com/?fe=1\&mimi2=1\&pact=9223685707\&utm\_camp\\ aign=*ad*+me+oh+my+graphics+35\_+off+sale+at+tko!\&utm\_medium\\ =customer+promotion\&utm\_source=*$\sim$*$\sim$*+tko+scraps+*$\sim$*$\sim$*\end{tabular} &
  115.9M \\ \hline
https://hotfrog.in/companies/hi-technical\_* & 115.9M                                        \\ \hline
\end{tabular}
\caption{Top 5 invalid `*` URLs with their reported number of mementos.}
\label{tab:asterisk}
\end{table}

During our analysis, we identified an issue involving duplicate TimeMaps within our dataset. Specifically, after converting our root URLs to SURT format, we found 355 URLs sharing duplicate SURTs. This duplication resulted from the SURT library's use of a regular expression that removes the \texttt{www} prefix from URI-Rs, leading to two scenarios. 

First, URLs containing multiple \texttt{www} subdomains were properly merged, which may have been the intended purpose of removing the \texttt{www} prefix from URI-Rs. For example, URLs such as \texttt{https://daily.co.jp/}, \\\texttt{https://www4.daily.co.jp/}, \texttt{https://www8.daily.co.jp/}, and \texttt{https://www9.daily.co.jp/} were all mapped to the same SURT, \texttt{jp,co,daily)/}.

Second scenario demonstrates how different URLs that included \texttt{www} in their hostnames were incorrectly canonicalized together, conflating distinct domains into a single SURT representation.
For example, URLs such as \texttt{https://www1355544.com/}, \texttt{https://www3288.com/}, \texttt{https://www504778.com/}, \texttt{https://www556798.com/}, and \texttt{https://www57912.com/} were all mapped to the same SURT, \texttt{com)/}. This incorrect canonicalization led to distinct domains being treated as a single entity. 

To mitigate the issue, we removed the affected URLs with improper \texttt{www} hostname canonicalization from our dataset. Additionally, we proposed an update to the regular expression used by the SURT library: the regex should only remove the \texttt{"www\textbackslash d*."} if at least two dots are present in the hostname. This adjustment ensures more accurate and distinct canonicalization of URLs. We submitted this proposal by opening an issue on GitHub.\footnote{\url{https://github.com/internetarchive/surt/issues/28##issue-1286134409}}

\subsection{Final Dataset}\label{subsec:finaldata}

\begin{figure}
  \centering
  \frame{\includegraphics[width=\linewidth]{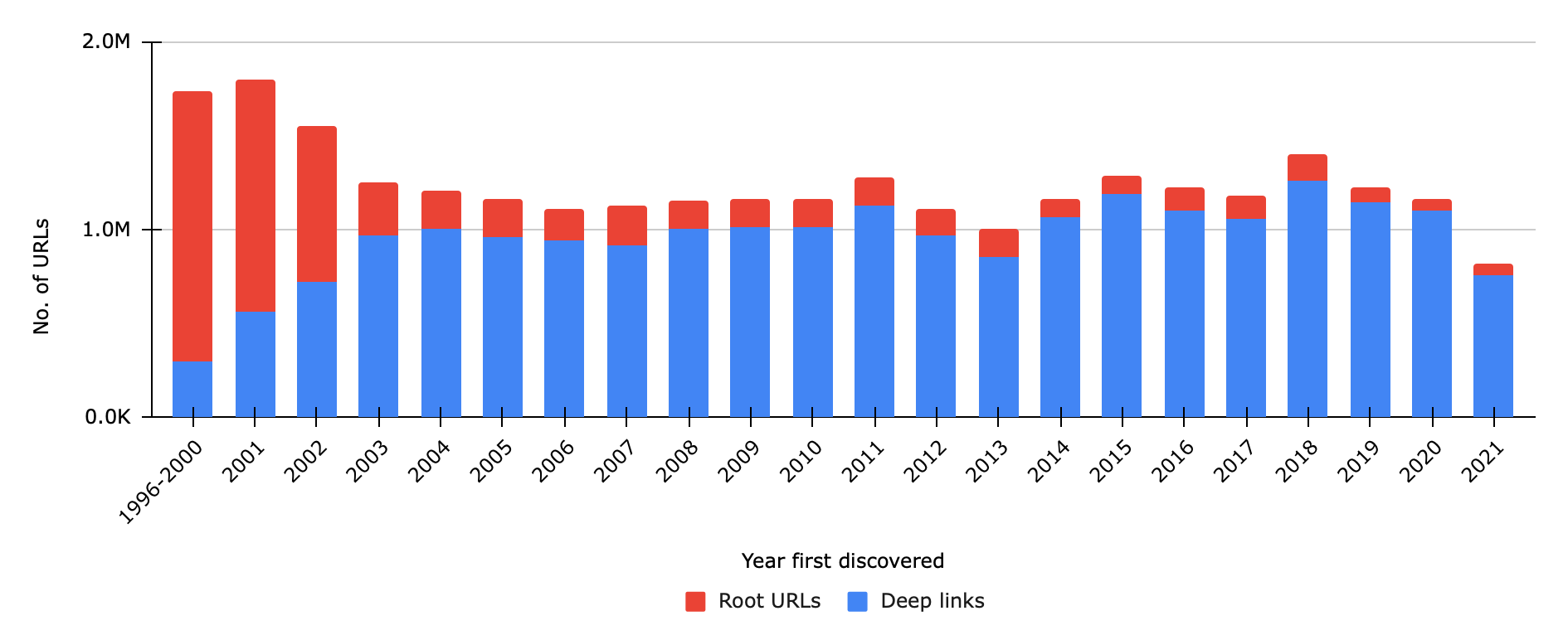}}
  \caption{Number of root URLs and deeplinks for each year in our final sample.}
  \label{img:firstarchivedfinalsample}
\end{figure}

Our final NYPW sample consists of TimeMaps for 27.3 million URLs, collected from 1996 to 2021. Figure~\ref{img:firstarchivedfinalsample} illustrates the distribution of root URLs and deep links across different years in our final dataset. Each bin contains more than one million URLs, except for 2021, which has fewer than one million URLs because the 2021 data covers only the first eight months, as the index file used was generated in August 2021. This dataset has also been analyzed and reported in various other studies~\cite{nypwfindingsblog2024,fil2024webisntforever, Garg2025nothere, garg2024someurls_ipres, alam2023nypw_dwebcamp, alam2024nypw_dwebcamp, garg2024poster_WAC2024, garg2023_lessonsampling}.

\begin{figure}
  \centering
  \frame{\includegraphics[width=\linewidth]{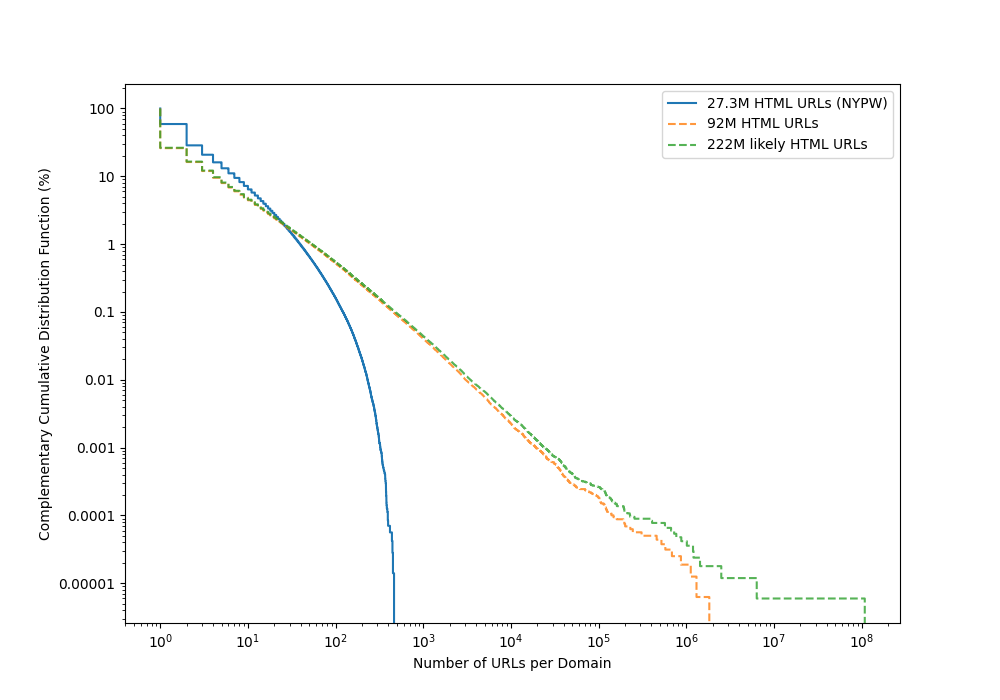}}
  \caption{Log-log plot of the Complementary Cumulative Distribution Function (CCDF) for the number of HTML URLs per domain, illustrating three datasets. The solid blue line represents a sample of 27.3 million HTML URLs (NYPW), the dashed orange line represents 92 million HTML URLs, and the dotted green line indicates 222 million likely HTML URLs. The x-axis shows the number of URLs per domain, while the y-axis represents the percentage of domains with URLs exceeding a given count.}
  \label{img:URLpdomain}
\end{figure}

The NYPW sample spans 7 million unique domains. Figure~\ref{img:URLpdomain} presents a log-log plot of the Complementary Cumulative Distribution Function (CCDF) for the number of HTML URLs per domain across three datasets, illustrating different stages of the downsampling process we used to refine web archival data. The green line (222M likely HTML URLs) represents the broadest dataset, derived from an initial filtering process based on file extensions to identify probable HTML pages. The orange line (92M HTML URLs) refines this set by filtering down to HTML pages with \texttt{text/html} MIME type. The blue line (27.3M HTML URLs - NYPW) is the final dataset, carefully downsampled to maintain meaningful domain representation while preventing a small number of popular domains from dominating the dataset. The early drop in the blue curve reflects the removal of domains with only a single URL. The dataset retains a heavy-tailed distribution, a common pattern in web data where a few domains contribute a large number of URLs while most contain only a few. The log-based downsampling approach ensures that no single domain dominates the dataset, balancing domain representation.

\begin{figure}
  \centering
  \frame{\includegraphics[width=\linewidth]{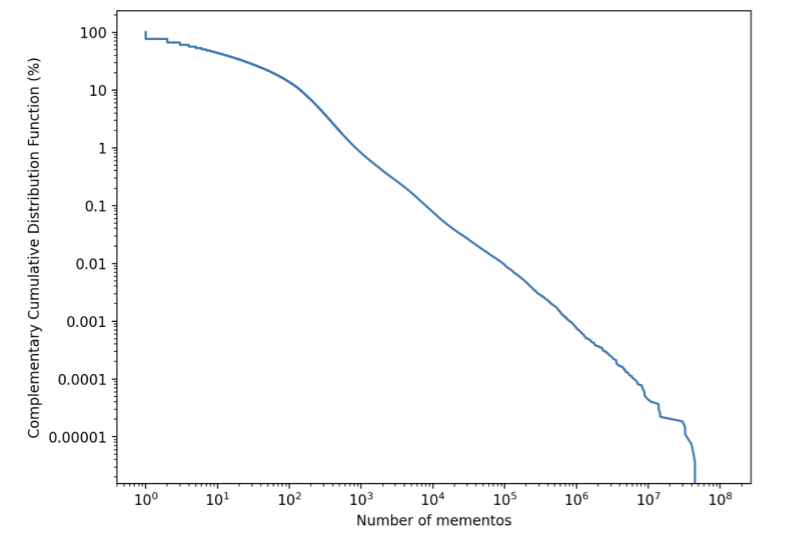}}
  \caption{Log-log plot of the CCDF representing the number of mementos per URL. The x-axis indicates the number of mementos, ranging from 1 to 100 million, while the y-axis shows the corresponding probability percentage. }
  \label{img:mementospURL}
\end{figure}

The NYPW sample includes 3.8 billion mementos spanning 27.3 million URLs. We analyzed the mementos associated with each URL in our dataset. Figure~\ref{img:mementospURL} presents a log-log plot of the CCDF for the number of mementos per URL. The plot exhibits a heavy-tailed distribution, where a small fraction of URLs have an extremely high number of mementos, while most have relatively few. The consistent slope in the central portion of the log-log plot suggests a power-law relationship, meaning that as the number of mementos increases, their frequency declines in a predictable manner. The initial plateau in the CCDF indicates that every URL in our dataset has at least one memento, with approximately 6 million URLs (22\% of the dataset) having just a single memento. This suggests that a substantial portion of the dataset consists of sparsely archived URLs, likely due to limited archival interest. As the number of mementos increases, the CCDF rapidly declines, confirming that URLs with high memento counts are increasingly rare. This pattern aligns with the Pareto principle, where a small percentage of items account for the majority of occurrences~\cite{pareto1896cours}. 

\section{Discussion}\label{sec:discussion}

It is important to note that despite our efforts to balance domain representation, other forms of bias may still be present in the data. Given the vastness and complexity of the web, constructing a truly representative dataset of the entire web is impossible. Instead, we relied on archived web data as our best available option, recognizing that it carries inherent biases based on what was preserved. The archivable web is inherently a subset of the entire web \cite{hallak2018privateweb}, shaped by factors such as crawler policies, technical constraints, and institutional priorities. Although our data set is not a perfect reflection of the web, it might be representative of the archived web itself, inheriting its tendencies, biases, and limitations. To mitigate some of these biases, we took deliberate steps to reduce the dominance of highly popular domains and limit the number of domains with only a single URL, which made up a significant portion of the dataset. Our goal was to ensure that each yearly sample contained approximately 1 million HTML URLs, providing a structured and scalable dataset for analysis rather than aiming for exhaustive web coverage.

One of the potential biases that may arise from using the archived web to curate a sample is the difference between how an archive crawler and a human user experience the web. Crawlers, unlike human users, are often subject to cloaking techniques~\cite{Wu2005cloaking, GoogleCloaking}, where websites serve different content based on the type of visitor. This means that what is captured in an archive may not always reflect what a typical user would have seen at the time. For instance, some websites block automated crawlers~\cite{Klein2020Persistence}. Others may serve decoy content to crawlers while displaying full content to human users, making it difficult to verify whether a page replayed from an archive is an authentic representation of what users originally encountered. Moreover, validating whether a web page replays well at this scale is a significant challenge~\cite{banos2015quantitative,brunelle2014mementodamage}. With 3.8 billion mementos in our dataset, de-referencing and fully rendering each one to assess replay accuracy is computationally expensive and time-consuming. Some pages may have captured only partial content, broken styles, or missing embedded resources, further complicating the process of ensuring data reliability. Similarly, extracting the content language of archived pages presents another difficulty. Given that our dataset contains billions of mementos spanning multiple languages and regions, automatically detecting and categorizing the language of every page is a non-trivial task. While HTML-based text extraction techniques exist, the sheer scale of the dataset makes language identification a resource-intensive process that cannot be easily applied to all pages. This limitation may introduce biases if certain languages are underrepresented. For instance, \citet{lulwah2017} found that the archiving rates of web pages vary significantly across languages, with English web pages being archived at a much higher rate (72.04\%) than Arabic (53\%), Danish (35.89\%), and Korean (32.81\%) pages. This discrepancy highlights how language representation in web archives is uneven, potentially leading to biases in language-based web studies.

Ultimately, these challenges highlight the inherent limitations of working with web archives at scale. While our dataset is shaped by archival biases rather than those of the live web, we made efforts to balance domain representation in our sample, reducing the over-sampling of heavily represented domains and minimizing the long tail of domains that only appear once (i.e., domains with only a single URL in the sample). Addressing the remaining challenges would require significant computational resources, improved archival methodologies, and new approaches for large-scale analysis. Despite these limitations, our dataset provides a structured and scalable collection of archived web pages, enabling further research on the historical web.

\section{Conclusions}\label{sec:conclusion}

In this paper, we documented our methodology for sampling archived web to curate a web sample of 27.3 million URLs and 3.8 billion archived pages spanning 26 years (1996–2021) from the Internet Archive’s Wayback Machine. The overarching goal of \textit{Not Your Parents’ Web} (NYPW) is to revisit fundamental questions about the publicly archivable web, particularly: “How long does a web page last?”

To explore this question, we developed a dataset by systematically sampling URLs from the Internet Archive’s ZipNum index file. We initially selected 285 million unique SURTs and conducted multiple refinement steps to produce a web sample that would yield approximately one million HTML URLs for each year of the Internet Archive.

We applied a series of filtering steps to extract and refine URLs from large-scale archival data, which involved:
\begin{itemize}
    \item Sampled URLs efficiently from the Internet Archive’s large-scale CDX data using the ZipNum index file.
    \item Removed likely URL aliases (e.g., session IDs, duplicate index file variants) to improve uniqueness.
    \item Identified and extracted likely HTML pages based on file extensions, reducing computational overhead.
    \item Verified that 86.12\% of the 107 million likely HTML URLs had a \texttt{text/html} MIME type through CDX API queries.
\end{itemize}

To account for temporal and domain-level biases, we adjusted our sampling methodology, including:
\begin{itemize}
    \item Determined the first archival year of each URL via CDX API queries.
    \item Upsampled early web pages (1996–2000) by extracting root URLs from deep links, addressing archival sparsity.
    \item Mitigated the over-representation of popular domains by initially excluding them, followed by a controlled reintegration process and logarithmic downsampling.
    \item Balanced URL distribution per domain, reducing dominance from frequently archived domains and ensuring broader domain coverage.
\end{itemize}

We retrieved complete TimeMaps for the final 27.3 million downsampled URLs using the CDX Pagination API. We identified a few issues and resolved inconsistencies in TimeMaps, including:
\begin{itemize}
    \item “warc/revisit” records, rehydrated using an LRU cache for efficiency.
    \item Invalid wildcard\texttt{*} URLs, removed due to misinterpretation by the CDX server.
    \item Canonicalization errors in the SURT library, identified and fixed through an implemented patch.
\end{itemize}

The resulting dataset, comprising 27.3 million URLs and 3.8 billion mementos from 1996 to 2021, provides a structured resource for examining web persistence~\cite{nypw_githubrepo}. While archived web data carries inherent biases, this dataset offers a foundation for further research into web evolution and longevity. The methods and findings from this study can inform future efforts in web archive analysis and dataset construction.

\begin{acks}
This work is supported in part by the Internet Archive and Filecoin Foundation \cite{nypwblog2021}.

\end{acks}

\bibliographystyle{ACM-Reference-Format}
\bibliography{ref}


\begin{thebibliography}{73}


\ifx \showCODEN    \undefined \def \showCODEN     #1{\unskip}     \fi
\ifx \showDOI      \undefined \def \showDOI       #1{#1}\fi
\ifx \showISBNx    \undefined \def \showISBNx     #1{\unskip}     \fi
\ifx \showISBNxiii \undefined \def \showISBNxiii  #1{\unskip}     \fi
\ifx \showISSN     \undefined \def \showISSN      #1{\unskip}     \fi
\ifx \showLCCN     \undefined \def \showLCCN      #1{\unskip}     \fi
\ifx \shownote     \undefined \def \shownote      #1{#1}          \fi
\ifx \showarticletitle \undefined \def \showarticletitle #1{#1}   \fi
\ifx \showURL      \undefined \def \showURL       {\relax}        \fi
\providecommand\bibfield[2]{#2}
\providecommand\bibinfo[2]{#2}
\providecommand\natexlab[1]{#1}
\providecommand\showeprint[2][]{arXiv:#2}

\bibitem[Adar et~al\mbox{.}(2009)]%
        {Eytan2009webchanges}
\bibfield{author}{\bibinfo{person}{Eytan Adar}, \bibinfo{person}{Jaime Teevan}, \bibinfo{person}{Susan~T. Dumais}, {and} \bibinfo{person}{Jonathan~L. Elsas}.} \bibinfo{year}{2009}\natexlab{}.
\newblock \showarticletitle{The web changes everything: Understanding the dynamics of web content}. In \bibinfo{booktitle}{\emph{Proceedings of the Second ACM International Conference on Web Search and Data Mining}} (Barcelona, Spain) \emph{(\bibinfo{series}{WSDM '09})}. \bibinfo{publisher}{Association for Computing Machinery}, \bibinfo{address}{New York, NY, USA}, \bibinfo{pages}{282–291}.
\newblock
\showISBNx{9781605583907}
\urldef\tempurl%
\url{https://doi.org/10.1145/1498759.1498837}
\showDOI{\tempurl}


\bibitem[Agata et~al\mbox{.}(2014)]%
        {agata2014lifespan}
\bibfield{author}{\bibinfo{person}{Teru Agata}, \bibinfo{person}{Yosuke Miyata}, \bibinfo{person}{Emi Ishita}, \bibinfo{person}{Atsushi Ikeuchi}, {and} \bibinfo{person}{Shuichi Ueda}.} \bibinfo{year}{2014}\natexlab{}.
\newblock \showarticletitle{Life Span of Web Pages: A Survey of 10 Million Pages Collected in 2001}. In \bibinfo{booktitle}{\emph{Proceedings of the 14th ACM/IEEE-CS Joint Conference on Digital Libraries (JCDL 2014)}}. \bibinfo{publisher}{IEEE}, \bibinfo{address}{London, UK}, \bibinfo{pages}{463--464}.
\newblock
\urldef\tempurl%
\url{https://doi.org/10.1109/JCDL.2014.6970226}
\showDOI{\tempurl}


\bibitem[Ainsworth et~al\mbox{.}(2011)]%
        {Ainsworth2011}
\bibfield{author}{\bibinfo{person}{Scott~G. Ainsworth}, \bibinfo{person}{Ahmed Alsum}, \bibinfo{person}{Hany SalahEldeen}, \bibinfo{person}{Michele~C. Weigle}, {and} \bibinfo{person}{Michael~L. Nelson}.} \bibinfo{year}{2011}\natexlab{}.
\newblock \showarticletitle{How much of the web is archived?}. In \bibinfo{booktitle}{\emph{Proceedings of the 11th Annual International ACM/IEEE Joint Conference on Digital Libraries}} (Ottawa, Ontario, Canada) \emph{(\bibinfo{series}{JCDL '11})}. \bibinfo{publisher}{Association for Computing Machinery}, \bibinfo{address}{New York, NY, USA}, \bibinfo{pages}{133–136}.
\newblock
\showISBNx{9781450307444}
\urldef\tempurl%
\url{https://doi.org/10.1145/1998076.1998100}
\showDOI{\tempurl}


\bibitem[Alam(2020)]%
        {alam2020mementomapthesis}
\bibfield{author}{\bibinfo{person}{Sawood Alam}.} \bibinfo{year}{2020}\natexlab{}.
\newblock \emph{\bibinfo{title}{MementoMap: A Web Archive Profiling Framework for Efficient Memento Routing}}.
\newblock PhD dissertation. \bibinfo{school}{Old Dominion University}, \bibinfo{address}{Norfolk, VA, USA}.
\newblock
\urldef\tempurl%
\url{https://doi.org/10.25777/5vnk-s536}
\showDOI{\tempurl}


\bibitem[Alam(2024)]%
        {alam_streamsampler}
\bibfield{author}{\bibinfo{person}{Sawood Alam}.} \bibinfo{year}{2024}\natexlab{}.
\newblock \bibinfo{title}{stream\_sampler.py — Reservoir sampling utility}.
\newblock \bibinfo{howpublished}{GitHub}.
\newblock
\urldef\tempurl%
\url{https://github.com/ibnesayeed/utils/blob/main/textfiles/stream_sampler.py}
\showURL{%
\tempurl}


\bibitem[Alam et~al\mbox{.}(2023a)]%
        {TrendMachine:2023}
\bibfield{author}{\bibinfo{person}{Sawood Alam}, \bibinfo{person}{Kritika Garg}, \bibinfo{person}{Michele~C. Weigle}, \bibinfo{person}{Michael~L. Nelson}, \bibinfo{person}{Mark Graham}, {and} \bibinfo{person}{Dietrich Ayala}.} \bibinfo{year}{2023}\natexlab{a}.
\newblock \showarticletitle{TrendMachine: A Temporal Webpage Resilience Portal}. In \bibinfo{booktitle}{\emph{Proceedings of the ACM/IEEE Joint Conference on Digital Libraries (JCDL)}}. \bibinfo{publisher}{IEEE Press}, \bibinfo{address}{Santa Fe, New Mexico, USA}, \bibinfo{pages}{93--97}.
\newblock
\urldef\tempurl%
\url{https://doi.org/10.1109/JCDL57899.2023.00023}
\showDOI{\tempurl}


\bibitem[Alam et~al\mbox{.}(2023b)]%
        {alam2023nypw_dwebcamp}
\bibfield{author}{\bibinfo{person}{Sawood Alam}, \bibinfo{person}{Mark Graham}, \bibinfo{person}{Kritika Garg}, \bibinfo{person}{Michele~C. Weigle}, \bibinfo{person}{Michael~L. Nelson}, {and} \bibinfo{person}{Dietrich Ayala}.} \bibinfo{year}{2023}\natexlab{b}.
\newblock \bibinfo{title}{Not Your Parents’ Web}.
\newblock \bibinfo{howpublished}{DWeb Camp 2023}.
\newblock
\urldef\tempurl%
\url{https://dwebcamp2023.sched.com/event/1NnA6/not-your-parents-web}
\showURL{%
\tempurl}


\bibitem[Alam et~al\mbox{.}(2024)]%
        {alam2024nypw_dwebcamp}
\bibfield{author}{\bibinfo{person}{Sawood Alam}, \bibinfo{person}{Mark Graham}, \bibinfo{person}{Kritika Garg}, \bibinfo{person}{Michele~C. Weigle}, \bibinfo{person}{Michael~L. Nelson}, {and} \bibinfo{person}{Dietrich Ayala}.} \bibinfo{year}{2024}\natexlab{}.
\newblock \bibinfo{title}{Some URLs Are Immortal, Most Are Ephemeral}.
\newblock \bibinfo{howpublished}{DWeb Camp 2024}.
\newblock
\urldef\tempurl%
\url{https://dwebcamp2024.sched.com/event/1hX2l/some-urls-are-immortal-most-are-ephemeral}
\showURL{%
\tempurl}


\bibitem[Alam and Nelson(2016)]%
        {alam2016memgator}
\bibfield{author}{\bibinfo{person}{Sawood Alam} {and} \bibinfo{person}{Michael~L. Nelson}.} \bibinfo{year}{2016}\natexlab{}.
\newblock \showarticletitle{MemGator --- A Portable Concurrent Memento Aggregator: Cross-Platform CLI and Server Binaries in Go}. In \bibinfo{booktitle}{\emph{Proceedings of the 2016 IEEE/ACM Joint Conference on Digital Libraries (JCDL)}}. \bibinfo{publisher}{IEEE}, \bibinfo{pages}{243--244}.
\newblock
\urldef\tempurl%
\url{https://doi.org/10.1145/2910896.2910916}
\showDOI{\tempurl}


\bibitem[Alam et~al\mbox{.}(2019)]%
        {alam2019mementomap}
\bibfield{author}{\bibinfo{person}{Sawood Alam}, \bibinfo{person}{Michele Weigle}, \bibinfo{person}{Michael Nelson}, \bibinfo{person}{Fernando Melo}, \bibinfo{person}{Daniel Bicho}, {and} \bibinfo{person}{Daniel Gomes}.} \bibinfo{year}{2019}\natexlab{}.
\newblock \showarticletitle{MementoMap Framework for Flexible and Adaptive Web Archive Profiling}. In \bibinfo{booktitle}{\emph{Proceedings of the 2019 ACM/IEEE Joint Conference on Digital Libraries (JCDL)}}. \bibinfo{publisher}{IEEE}, \bibinfo{pages}{172--181}.
\newblock
\urldef\tempurl%
\url{https://doi.org/10.1109/JCDL.2019.00033}
\showDOI{\tempurl}


\bibitem[Alkwai et~al\mbox{.}(2017)]%
        {lulwah2017}
\bibfield{author}{\bibinfo{person}{Lulwah~M. Alkwai}, \bibinfo{person}{Michael~L. Nelson}, {and} \bibinfo{person}{Michele~C. Weigle}.} \bibinfo{year}{2017}\natexlab{}.
\newblock \showarticletitle{Comparing the Archival Rate of Arabic, English, Danish, and Korean Language Web Pages}.
\newblock \bibinfo{journal}{\emph{ACM Transactions on Information Systems}} \bibinfo{volume}{36}, \bibinfo{number}{1}, Article \bibinfo{articleno}{1} (\bibinfo{date}{June} \bibinfo{year}{2017}), \bibinfo{numpages}{34}~pages.
\newblock
\showISSN{1046-8188}
\urldef\tempurl%
\url{https://doi.org/10.1145/3041656}
\showDOI{\tempurl}


\bibitem[AlSum et~al\mbox{.}(2014)]%
        {AlSum2014}
\bibfield{author}{\bibinfo{person}{Ahmed AlSum}, \bibinfo{person}{Michele~C. Weigle}, \bibinfo{person}{Michael~L. Nelson}, {and} \bibinfo{person}{Herbert Van~de Sompel}.} \bibinfo{year}{2014}\natexlab{}.
\newblock \showarticletitle{Profiling Web Archive Coverage for Top-Level Domain and Content Language}.
\newblock \bibinfo{journal}{\emph{International Journal on Digital Libraries}} \bibinfo{volume}{14}, \bibinfo{number}{3-4} (\bibinfo{date}{August} \bibinfo{year}{2014}), \bibinfo{pages}{149--166}.
\newblock
\urldef\tempurl%
\url{https://doi.org/10.1007/s00799-014-0118-y}
\showDOI{\tempurl}


\bibitem[Archive(2024a)]%
        {cdx_file_format}
\bibfield{author}{\bibinfo{person}{Internet Archive}.} \bibinfo{year}{2024}\natexlab{a}.
\newblock \bibinfo{title}{CDX File Format Specification}.
\newblock \bibinfo{howpublished}{\url{https://archive.org/web/researcher/cdx_file_format.php}}.
\newblock


\bibitem[Archive(2024b)]%
        {surt_repository}
\bibfield{author}{\bibinfo{person}{Internet Archive}.} \bibinfo{year}{2024}\natexlab{b}.
\newblock \bibinfo{title}{SURT (Sort-friendly URI Reordering Transform)}.
\newblock \bibinfo{howpublished}{\url{https://github.com/internetarchive/surt}}.
\newblock


\bibitem[Archive(2024c)]%
        {wayback_cdx_server}
\bibfield{author}{\bibinfo{person}{Internet Archive}.} \bibinfo{year}{2024}\natexlab{c}.
\newblock \bibinfo{title}{Wayback CDX Server API Documentation}.
\newblock \bibinfo{howpublished}{\url{https://github.com/internetarchive/wayback/blob/master/wayback-cdx-server/README.md}}.
\newblock


\bibitem[Banos and Manolopoulos(2015)]%
        {banos2015quantitative}
\bibfield{author}{\bibinfo{person}{Vangelis Banos} {and} \bibinfo{person}{Yannis Manolopoulos}.} \bibinfo{year}{2015}\natexlab{}.
\newblock \showarticletitle{A Quantitative Approach to Evaluate Website Archivability Using the CLEAR+ Method}.
\newblock \bibinfo{journal}{\emph{International Journal on Digital Libraries}} (\bibinfo{year}{2015}), \bibinfo{pages}{1--23}.
\newblock
\urldef\tempurl%
\url{https://doi.org/10.1007/s00799-015-0144-4}
\showDOI{\tempurl}


\bibitem[Bar-Yossef et~al\mbox{.}(2009)]%
        {Bar-Yossef2009}
\bibfield{author}{\bibinfo{person}{Ziv Bar-Yossef}, \bibinfo{person}{Idit Keidar}, {and} \bibinfo{person}{Uri Schonfeld}.} \bibinfo{year}{2009}\natexlab{}.
\newblock \showarticletitle{Do not crawl in the DUST: Different URLs with similar text}.
\newblock \bibinfo{journal}{\emph{ACM Transactions on the Web}} \bibinfo{volume}{3}, \bibinfo{number}{1}, Article \bibinfo{articleno}{3} (\bibinfo{date}{Jan.} \bibinfo{year}{2009}), \bibinfo{numpages}{31}~pages.
\newblock
\showISSN{1559-1131}
\urldef\tempurl%
\url{https://doi.org/10.1145/1462148.1462151}
\showDOI{\tempurl}


\bibitem[Berners-Lee et~al\mbox{.}(1994)]%
        {BernersLee1994}
\bibfield{author}{\bibinfo{person}{Tim Berners-Lee}, \bibinfo{person}{Robert Cailliau}, \bibinfo{person}{Ari Luotonen}, \bibinfo{person}{Henrik~Frystyk Nielsen}, {and} \bibinfo{person}{Arthur Secret}.} \bibinfo{year}{1994}\natexlab{}.
\newblock \showarticletitle{The World-Wide Web}.
\newblock \bibinfo{journal}{\emph{Commun. ACM}} \bibinfo{volume}{37}, \bibinfo{number}{8} (\bibinfo{date}{August} \bibinfo{year}{1994}), \bibinfo{pages}{76–82}.
\newblock
\showISSN{0001-0782}
\urldef\tempurl%
\url{https://doi.org/10.1145/179606.179671}
\showDOI{\tempurl}


\bibitem[Berners-Lee et~al\mbox{.}(2005)]%
        {rfc3986}
\bibfield{author}{\bibinfo{person}{Tim Berners-Lee}, \bibinfo{person}{Roy~T. Fielding}, {and} \bibinfo{person}{Larry Masinter}.} \bibinfo{year}{2005}\natexlab{}.
\newblock \bibinfo{title}{RFC 3986 - Uniform Resource Identifier (URI): Generic Syntax}.
\newblock \bibinfo{howpublished}{Internet Engineering Task Force (IETF)}.
\newblock
\urldef\tempurl%
\url{https://www.rfc-editor.org/rfc/rfc3986}
\showURL{%
\tempurl}


\bibitem[Brewington and Cybenko(2000)]%
        {BREWINGTON2000257}
\bibfield{author}{\bibinfo{person}{Brian~E. Brewington} {and} \bibinfo{person}{George Cybenko}.} \bibinfo{year}{2000}\natexlab{}.
\newblock \showarticletitle{How dynamic is the Web?}
\newblock \bibinfo{journal}{\emph{Computer Networks}} \bibinfo{volume}{33}, \bibinfo{number}{1} (\bibinfo{year}{2000}), \bibinfo{pages}{257--276}.
\newblock
\showISSN{1389-1286}
\urldef\tempurl%
\url{https://doi.org/10.1016/S1389-1286(00)00045-1}
\showDOI{\tempurl}


\bibitem[Brin and Page(1998)]%
        {brin1998anatomy}
\bibfield{author}{\bibinfo{person}{Sergey Brin} {and} \bibinfo{person}{Lawrence Page}.} \bibinfo{year}{1998}\natexlab{}.
\newblock \showarticletitle{The anatomy of a large-scale hypertextual web search engine}.
\newblock \bibinfo{journal}{\emph{Computer Networks and ISDN Systems}} \bibinfo{volume}{30}, \bibinfo{number}{1-7} (\bibinfo{year}{1998}), \bibinfo{pages}{107--117}.
\newblock
\urldef\tempurl%
\url{https://doi.org/10.1016/S0169-7552(98)00110-X}
\showDOI{\tempurl}


\bibitem[Brunelle et~al\mbox{.}(2014)]%
        {brunelle2014mementodamage}
\bibfield{author}{\bibinfo{person}{Justin~F. Brunelle}, \bibinfo{person}{Mat Kelly}, \bibinfo{person}{Hany SalahEldeen}, \bibinfo{person}{Michele~C. Weigle}, {and} \bibinfo{person}{Michael~L. Nelson}.} \bibinfo{year}{2014}\natexlab{}.
\newblock \showarticletitle{Not All Mementos Are Created Equal: Measuring the Impact of Missing Resources}. In \bibinfo{booktitle}{\emph{Proceedings of the IEEE/ACM Joint Conference on Digital Libraries (JCDL)}}. \bibinfo{publisher}{IEEE}, \bibinfo{pages}{321--330}.
\newblock
\urldef\tempurl%
\url{https://doi.org/10.1109/JCDL.2014.6970187}
\showDOI{\tempurl}


\bibitem[Chapekis et~al\mbox{.}(2024)]%
        {pew2024}
\bibfield{author}{\bibinfo{person}{Athena Chapekis}, \bibinfo{person}{Samuel Bestvater}, \bibinfo{person}{Emma Remy}, {and} \bibinfo{person}{Gonzalo Rivero}.} \bibinfo{year}{2024}\natexlab{}.
\newblock \bibinfo{title}{When Online Content Disappears}.
\newblock
\newblock
\urldef\tempurl%
\url{https://www.pewresearch.org/data-labs/2024/05/17/when-online-content-disappears/}
\showURL{%
\tempurl}


\bibitem[Cho and Garcia-Molina(2000)]%
        {cho2000webevolution}
\bibfield{author}{\bibinfo{person}{Junghoo Cho} {and} \bibinfo{person}{Hector Garcia-Molina}.} \bibinfo{year}{2000}\natexlab{}.
\newblock \showarticletitle{The Evolution of the Web and Implications for an Incremental Crawler}. In \bibinfo{booktitle}{\emph{Proceedings of the 26th International Conference on Very Large Data Bases}} \emph{(\bibinfo{series}{VLDB '00})}. \bibinfo{publisher}{Morgan Kaufmann Publishers Inc.}, \bibinfo{address}{San Francisco, CA, USA}, \bibinfo{pages}{200–209}.
\newblock
\showISBNx{1558607153}


\bibitem[contributors(2024)]%
        {wikipedia_cache_replacement}
\bibfield{author}{\bibinfo{person}{Wikipedia contributors}.} \bibinfo{year}{2024}\natexlab{}.
\newblock \bibinfo{title}{Cache replacement policies -- Wikipedia, The Free Encyclopedia}.
\newblock \bibinfo{howpublished}{\url{https://en.wikipedia.org/wiki/Cache_replacement_policies}}.
\newblock


\bibitem[de~Sompel et~al\mbox{.}(2013)]%
        {rfc7089}
\bibfield{author}{\bibinfo{person}{Herbert~Van de Sompel}, \bibinfo{person}{Michael~L. Nelson}, \bibinfo{person}{Robert Sanderson}, \bibinfo{person}{Lyudmila Balakireva}, \bibinfo{person}{Scott Ainsworth}, {and} \bibinfo{person}{Harihar Shankar}.} \bibinfo{year}{2013}\natexlab{}.
\newblock \bibinfo{title}{RFC 7089 - HTTP Framework for Time-Based Access to Resource States -- Memento}.
\newblock \bibinfo{howpublished}{Internet Engineering Task Force (IETF)}.
\newblock
\urldef\tempurl%
\url{https://tools.ietf.org/html/rfc7089}
\showURL{%
\tempurl}


\bibitem[Eastlake and Panitz(1999)]%
        {rfc2606}
\bibfield{author}{\bibinfo{person}{Donald Eastlake} {and} \bibinfo{person}{Al Panitz}.} \bibinfo{year}{1999}\natexlab{}.
\newblock \bibinfo{title}{RFC 2606 - Reserved Top Level DNS Names}.
\newblock \bibinfo{howpublished}{Internet Engineering Task Force (IETF)}.
\newblock
\urldef\tempurl%
\url{https://www.rfc-editor.org/rfc/rfc2606}
\showURL{%
\tempurl}


\bibitem[Fetterly et~al\mbox{.}(2003)]%
        {Fetterly2003}
\bibfield{author}{\bibinfo{person}{Dennis Fetterly}, \bibinfo{person}{Mark Manasse}, \bibinfo{person}{Marc Najork}, {and} \bibinfo{person}{Janet Wiener}.} \bibinfo{year}{2003}\natexlab{}.
\newblock \showarticletitle{A large-scale study of the evolution of web pages}. In \bibinfo{booktitle}{\emph{Proceedings of the 12th International Conference on World Wide Web}} (Budapest, Hungary) \emph{(\bibinfo{series}{WWW '03})}. \bibinfo{publisher}{Association for Computing Machinery}, \bibinfo{address}{New York, USA}, \bibinfo{pages}{669–678}.
\newblock
\showISBNx{1581136803}
\urldef\tempurl%
\url{https://doi.org/10.1145/775152.775246}
\showDOI{\tempurl}


\bibitem[{Filecoin Foundation}(2024)]%
        {fil2024webisntforever}
\bibfield{author}{\bibinfo{person}{{Filecoin Foundation}}.} \bibinfo{year}{2024}\natexlab{}.
\newblock \bibinfo{title}{The Web Isn’t Forever: New Research Findings from “Not Your Parents’ Web” Project}.
\newblock \bibinfo{howpublished}{\url{https://fil.org/blog/the-web-isn-t-forever-new-research-findings-from-not-your-parents-web-project}}.
\newblock


\bibitem[Garg et~al\mbox{.}(2024a)]%
        {garg2024someurls_ipres}
\bibfield{author}{\bibinfo{person}{Kritika Garg}, \bibinfo{person}{Sawood Alam}, \bibinfo{person}{Dietrich Ayala}, \bibinfo{person}{Mark Graham}, \bibinfo{person}{Michele~C. Weigle}, {and} \bibinfo{person}{Michael~L. Nelson}.} \bibinfo{year}{2024}\natexlab{a}.
\newblock \bibinfo{title}{Some URLs Are Immortal, Most Are Ephemeral}.
\newblock \bibinfo{howpublished}{Poster, iPRES 2024; published via Zenodo}.
\newblock
\urldef\tempurl%
\url{https://doi.org/10.5281/zenodo.13687116}
\showDOI{\tempurl}


\bibitem[Garg et~al\mbox{.}(2024b)]%
        {garg2024poster_WAC2024}
\bibfield{author}{\bibinfo{person}{Kritika Garg}, \bibinfo{person}{Sawood Alam}, \bibinfo{person}{Dietrich Ayala}, \bibinfo{person}{Michele~C. Weigle}, {and} \bibinfo{person}{Michael~L. Nelson}.} \bibinfo{year}{2024}\natexlab{b}.
\newblock \bibinfo{title}{{Some URLs Are Immortal, Most Are Ephemeral} (Poster \#203)}.
\newblock \bibinfo{howpublished}{{Poster session at the IIPC GA \& Web Archiving Conference 2024}}.
\newblock
\urldef\tempurl%
\url{https://netpreserve.org/ga2024/abstracts/#poster_203}
\showURL{%
\tempurl}
\newblock
\shownote{Poster \#203}.


\bibitem[Garg et~al\mbox{.}(2025)]%
        {Garg2025nothere}
\bibfield{author}{\bibinfo{person}{Kritika Garg}, \bibinfo{person}{Sawood Alam}, \bibinfo{person}{Dietrich Ayala}, \bibinfo{person}{Michele~C. Weigle}, {and} \bibinfo{person}{Michael~L. Nelson}.} \bibinfo{year}{2025}\natexlab{}.
\newblock \showarticletitle{Not Here, Go There: Analyzing Redirection Patterns on the Web}. In \bibinfo{booktitle}{\emph{Proceedings of the 17th ACM Web Science Conference}} \emph{(\bibinfo{series}{WebSci '25})}. \bibinfo{publisher}{Association for Computing Machinery}, \bibinfo{address}{New York, NY, USA}, \bibinfo{pages}{249--260}.
\newblock
\urldef\tempurl%
\url{https://doi.org/10.1145/3717867.3717925}
\showDOI{\tempurl}


\bibitem[Garg et~al\mbox{.}(2023)]%
        {garg2023_lessonsampling}
\bibfield{author}{\bibinfo{person}{Kritika Garg}, \bibinfo{person}{Sawood Alam}, \bibinfo{person}{Michele Weigle}, \bibinfo{person}{Michael Nelson}, \bibinfo{person}{Corentin Barreau}, {and} \bibinfo{person}{Mark Graham}.} \bibinfo{year}{2023}\natexlab{}.
\newblock \bibinfo{title}{{Lessons Learned From the Longitudinal Sampling of a Large Web Archive}}.
\newblock \bibinfo{howpublished}{UNT Digital Library}.
\newblock
\urldef\tempurl%
\url{https://digital.library.unt.edu/ark:/67531/metadc2143930/}
\showURL{%
\tempurl}


\bibitem[Garg et~al\mbox{.}(2021)]%
        {Garg2021Twitter}
\bibfield{author}{\bibinfo{person}{Kritika Garg}, \bibinfo{person}{Himarsha~R. Jayanetti}, \bibinfo{person}{Sawood Alam}, \bibinfo{person}{Michele~C. Weigle}, {and} \bibinfo{person}{Michael~L. Nelson}.} \bibinfo{year}{2021}\natexlab{}.
\newblock \showarticletitle{Replaying Archived Twitter: When your bird is broken, will it bring you down?}. In \bibinfo{booktitle}{\emph{Proceedings of the 2021 ACM/IEEE Joint Conference on Digital Libraries (JCDL)}}. \bibinfo{pages}{160--169}.
\newblock
\urldef\tempurl%
\url{https://doi.org/10.1109/JCDL52503.2021.00028}
\showDOI{\tempurl}


\bibitem[Garg et~al\mbox{.}(2024c)]%
        {garg2024twitterIJDL}
\bibfield{author}{\bibinfo{person}{Kritika Garg}, \bibinfo{person}{Himarsha~R. Jayanetti}, \bibinfo{person}{Sawood Alam}, \bibinfo{person}{Michele~C. Weigle}, {and} \bibinfo{person}{Michael~L. Nelson}.} \bibinfo{year}{2024}\natexlab{c}.
\newblock \showarticletitle{Challenges in Replaying Archived Twitter Pages}.
\newblock \bibinfo{journal}{\emph{International Journal on Digital Libraries}} \bibinfo{volume}{25}, \bibinfo{number}{2} (\bibinfo{year}{2024}), \bibinfo{pages}{217--236}.
\newblock
\urldef\tempurl%
\url{https://doi.org/10.1007/s00799-023-00379-w}
\showDOI{\tempurl}


\bibitem[Google(nd)]%
        {GoogleCloaking}
\bibfield{author}{\bibinfo{person}{Google}.} \bibinfo{year}{n.d.}\natexlab{}.
\newblock \bibinfo{title}{Cloaking - Search Essentials}.
\newblock
\newblock
\urldef\tempurl%
\url{https://developers.google.com/search/docs/essentials/spam-policies#cloaking}
\showURL{%
\tempurl}


\bibitem[Graham(2019)]%
        {Graham2019}
\bibfield{author}{\bibinfo{person}{Mark Graham}.} \bibinfo{year}{2019}\natexlab{}.
\newblock \bibinfo{title}{The Wayback Machine’s Save Page Now is New and Improved!}
\newblock
\newblock
\urldef\tempurl%
\url{https://blog.archive.org/2019/10/23/the-wayback-machines-save-page-now-is-new-and-improved/}
\showURL{%
\tempurl}


\bibitem[Hall and Tiropanis(2012)]%
        {WebScience2012}
\bibfield{author}{\bibinfo{person}{Wendy Hall} {and} \bibinfo{person}{Thanassis Tiropanis}.} \bibinfo{year}{2012}\natexlab{}.
\newblock \showarticletitle{Web evolution and Web Science}.
\newblock \bibinfo{journal}{\emph{Computer Networks}} \bibinfo{volume}{56}, \bibinfo{number}{18} (\bibinfo{year}{2012}), \bibinfo{pages}{3859--3865}.
\newblock
\showISSN{1389-1286}
\urldef\tempurl%
\url{https://doi.org/10.1016/j.comnet.2012.10.004}
\showDOI{\tempurl}


\bibitem[Hallak(2018)]%
        {hallak2018privateweb}
\bibfield{author}{\bibinfo{person}{Hussam Hallak}.} \bibinfo{year}{2018}\natexlab{}.
\newblock \bibinfo{title}{Why We Need Private Web Archives: Almost Two-Thirds of Web Traffic IS NOT Publicly Archivable}.
\newblock \bibinfo{howpublished}{\url{https://ws-dl.blogspot.com/2018/07/2018-07-18-why-we-need-private-web.html}}.
\newblock


\bibitem[Holzmann et~al\mbox{.}(2016)]%
        {holzmann2016dawn}
\bibfield{author}{\bibinfo{person}{Helge Holzmann}, \bibinfo{person}{Wolfgang Nejdl}, {and} \bibinfo{person}{Avishek Anand}.} \bibinfo{year}{2016}\natexlab{}.
\newblock \showarticletitle{The Dawn of Today's Popular Domains: A Study of the Archived German Web over 18 Years}. In \bibinfo{booktitle}{\emph{Proceedings of the 16th ACM/IEEE-CS Joint Conference on Digital Libraries (JCDL 2016)}}. \bibinfo{publisher}{ACM}, \bibinfo{address}{New Jersey, Newark, USA}, \bibinfo{pages}{73--82}.
\newblock
\urldef\tempurl%
\url{https://doi.org/10.1145/2910896.2910901}
\showDOI{\tempurl}


\bibitem[IIPC(2024)]%
        {warc_specification_1_1}
\bibfield{author}{\bibinfo{person}{IIPC}.} \bibinfo{year}{2024}\natexlab{}.
\newblock \bibinfo{title}{WARC 1.1 Specification - Revisit Record}.
\newblock \bibinfo{howpublished}{\url{https://iipc.github.io/warc-specifications/specifications/warc-format/warc-1.1/##revisit}}.
\newblock


\bibitem[Jacobs and Walsh(2004)]%
        {w3c_webarch}
\bibfield{author}{\bibinfo{person}{Ian Jacobs} {and} \bibinfo{person}{Norman Walsh}.} \bibinfo{year}{2004}\natexlab{}.
\newblock \bibinfo{title}{Architecture of the World Wide Web, Volume One}.
\newblock \bibinfo{howpublished}{World Wide Web Consortium (W3C)}.
\newblock
\urldef\tempurl%
\url{https://www.w3.org/TR/webarch/}
\showURL{%
\tempurl}


\bibitem[Jayawardana et~al\mbox{.}(2020)]%
        {Jayawardana2020}
\bibfield{author}{\bibinfo{person}{Yasith Jayawardana}, \bibinfo{person}{Alexander~C. Nwala}, \bibinfo{person}{Gavindya Jayawardena}, \bibinfo{person}{Jian Wu}, \bibinfo{person}{Sampath Jayarathna}, \bibinfo{person}{Michael~L. Nelson}, {and} \bibinfo{person}{C.~Lee Giles}.} \bibinfo{year}{2020}\natexlab{}.
\newblock \showarticletitle{Modeling Updates of Scholarly Webpages Using Archived Data}. In \bibinfo{booktitle}{\emph{2020 IEEE International Conference on Big Data (Big Data)}}. \bibinfo{publisher}{IEEE}, \bibinfo{pages}{1868--1877}.
\newblock
\urldef\tempurl%
\url{https://doi.org/10.1109/BigData50022.2020.9377796}
\showDOI{\tempurl}


\bibitem[Jones et~al\mbox{.}(2016)]%
        {jones2016scholarly}
\bibfield{author}{\bibinfo{person}{Shawn~M. Jones}, \bibinfo{person}{Herbert~Van de Sompel}, \bibinfo{person}{Harihar Shankar}, \bibinfo{person}{Martin Klein}, \bibinfo{person}{Richard Tobin}, {and} \bibinfo{person}{Claire Grover}.} \bibinfo{year}{2016}\natexlab{}.
\newblock \showarticletitle{Scholarly Context Adrift: Three out of Four URI References Lead to Changed Content}.
\newblock \bibinfo{journal}{\emph{PLOS ONE}} \bibinfo{volume}{11}, \bibinfo{number}{12} (\bibinfo{year}{2016}), \bibinfo{pages}{e0167475}.
\newblock
\urldef\tempurl%
\url{https://doi.org/10.1371/journal.pone.0167475}
\showDOI{\tempurl}


\bibitem[Jones et~al\mbox{.}(2021)]%
        {memento:springerbook}
\bibfield{author}{\bibinfo{person}{Shawn~M. Jones}, \bibinfo{person}{Martin Klein}, \bibinfo{person}{Herbert {Van de Sompel}}, \bibinfo{person}{Michael~L. Nelson}, {and} \bibinfo{person}{Michele~C. Weigle}.} \bibinfo{year}{2021}\natexlab{}.
\newblock \showarticletitle{Interoperability for Accessing Versions of Web Resources with the {M}emento Protocol}.
\newblock In \bibinfo{booktitle}{\emph{The Past Web: Exploring Web Archives}}. \bibinfo{publisher}{Springer International Publishing}.
\newblock
\showISBNx{978-3-030-63290-8}


\bibitem[Kahle(2019)]%
        {kahle2019tweet}
\bibfield{author}{\bibinfo{person}{Brewster Kahle}.} \bibinfo{year}{2019}\natexlab{}.
\newblock \bibinfo{title}{The Internet Archive is working to provide free and open access to all the world’s knowledge.}
\newblock \bibinfo{howpublished}{Twitter}.
\newblock
\urldef\tempurl%
\url{https://x.com/brewster_kahle/status/1118172506777509890}
\showURL{%
\tempurl}
\newblock
\shownote{Accessed: 2024-10-09}.


\bibitem[Kelly et~al\mbox{.}(2017)]%
        {kelly2017impact}
\bibfield{author}{\bibinfo{person}{Mat Kelly}, \bibinfo{person}{Lulwah~M. Alkwai}, \bibinfo{person}{Sawood Alam}, \bibinfo{person}{Michael~L. Nelson}, \bibinfo{person}{Michele~C. Weigle}, {and} \bibinfo{person}{Herbert~Van de Sompel}.} \bibinfo{year}{2017}\natexlab{}.
\newblock \showarticletitle{Impact of URI Canonicalization on Memento Count}. In \bibinfo{booktitle}{\emph{Proceedings of the 17th ACM/IEEE Joint Conference on Digital Libraries (JCDL)}} (Toronto, Ontario, Canada) \emph{(\bibinfo{series}{JCDL '17})}. \bibinfo{publisher}{IEEE}, \bibinfo{pages}{303--304}.
\newblock
\urldef\tempurl%
\url{https://doi.org/10.1109/JCDL.2017.7991597}
\showDOI{\tempurl}


\bibitem[Klein and Balakireva(2020)]%
        {Klein2020Persistence}
\bibfield{author}{\bibinfo{person}{Martin Klein} {and} \bibinfo{person}{Lyudmila Balakireva}.} \bibinfo{year}{2020}\natexlab{}.
\newblock \showarticletitle{On the Persistence of Persistent Identifiers of the Scholarly Web}. In \bibinfo{booktitle}{\emph{Digital Libraries for Open Knowledge}}. \bibinfo{publisher}{Springer International Publishing}, \bibinfo{address}{Cham}, \bibinfo{pages}{102--115}.
\newblock
\showISBNx{978-3-030-54956-5}
\urldef\tempurl%
\url{https://doi.org/10.1007/978-3-030-54956-5_8}
\showDOI{\tempurl}


\bibitem[Klein et~al\mbox{.}(2014)]%
        {klein2014scholarly}
\bibfield{author}{\bibinfo{person}{Martin Klein}, \bibinfo{person}{Herbert~Van de Sompel}, \bibinfo{person}{Robert Sanderson}, \bibinfo{person}{Harihar Shankar}, \bibinfo{person}{Lyudmila Balakireva}, \bibinfo{person}{Ke Zhou}, {and} \bibinfo{person}{Richard Tobin}.} \bibinfo{year}{2014}\natexlab{}.
\newblock \showarticletitle{Scholarly Context Not Found: One in Five Articles Suffers from Reference Rot}.
\newblock \bibinfo{journal}{\emph{PLOS ONE}} \bibinfo{volume}{9}, \bibinfo{number}{12} (\bibinfo{year}{2014}), \bibinfo{pages}{e115253}.
\newblock
\urldef\tempurl%
\url{https://doi.org/10.1371/journal.pone.0115253}
\showDOI{\tempurl}


\bibitem[Koehler(1999)]%
        {koehler1999webpageconstancy}
\bibfield{author}{\bibinfo{person}{Wallace Koehler}.} \bibinfo{year}{1999}\natexlab{}.
\newblock \showarticletitle{An Analysis of Web Page and Web Site Constancy and Permanence}.
\newblock \bibinfo{journal}{\emph{Journal of the American Society for Information Science}} \bibinfo{volume}{50}, \bibinfo{number}{2} (\bibinfo{year}{1999}), \bibinfo{pages}{162--180}.
\newblock
\urldef\tempurl%
\url{https://doi.org/10.1002/(SICI)1097-4571(1999)50:2<162::AID-ASI7>3.0.CO;2-B}
\showDOI{\tempurl}


\bibitem[Koehler(2002)]%
        {Koehler2002}
\bibfield{author}{\bibinfo{person}{Wallace Koehler}.} \bibinfo{year}{2002}\natexlab{}.
\newblock \showarticletitle{Web Page Change and Persistence: A Four-Year Longitudinal Study}.
\newblock \bibinfo{journal}{\emph{Journal of the American Society for Information Science and Technology}} \bibinfo{volume}{53}, \bibinfo{number}{2} (\bibinfo{year}{2002}), \bibinfo{pages}{162--171}.
\newblock
\urldef\tempurl%
\url{https://doi.org/10.1002/asi.10018}
\showDOI{\tempurl}


\bibitem[Kreymer(2024a)]%
        {pywb_indexing}
\bibfield{author}{\bibinfo{person}{Ilya Kreymer}.} \bibinfo{year}{2024}\natexlab{a}.
\newblock \bibinfo{title}{Indexing in PyWB}.
\newblock \bibinfo{howpublished}{\url{https://pywb.readthedocs.io/en/latest/manual/indexing.html}}.
\newblock


\bibitem[Kreymer(2024b)]%
        {webarchive_indexing}
\bibfield{author}{\bibinfo{person}{Ilya Kreymer}.} \bibinfo{year}{2024}\natexlab{b}.
\newblock \bibinfo{title}{Web Archive Indexing Tools}.
\newblock \bibinfo{howpublished}{\url{https://github.com/ikreymer/webarchive-indexing}}.
\newblock


\bibitem[Major(2021)]%
        {Major2021WebEphemera}
\bibfield{author}{\bibinfo{person}{Daniela Major}.} \bibinfo{year}{2021}\natexlab{}.
\newblock \showarticletitle{The Problem of Web Ephemera}.
\newblock In \bibinfo{booktitle}{\emph{The Past Web: Exploring Web Archives}}. \bibinfo{publisher}{Springer International Publishing}, \bibinfo{address}{Cham}, \bibinfo{pages}{5--10}.
\newblock
\showISBNx{978-3-030-63291-5}
\urldef\tempurl%
\url{https://doi.org/10.1007/978-3-030-63291-5_1}
\showDOI{\tempurl}


\bibitem[Major and Gomes(2021)]%
        {Major2021CollectiveMemory}
\bibfield{author}{\bibinfo{person}{Daniela Major} {and} \bibinfo{person}{Daniel Gomes}.} \bibinfo{year}{2021}\natexlab{}.
\newblock \showarticletitle{Web Archives Preserve Our Digital Collective Memory}.
\newblock In \bibinfo{booktitle}{\emph{The Past Web: Exploring Web Archives}}. \bibinfo{publisher}{Springer International Publishing}, \bibinfo{address}{Cham}, \bibinfo{pages}{11--19}.
\newblock
\showISBNx{978-3-030-63291-5}
\urldef\tempurl%
\url{https://doi.org/10.1007/978-3-030-63291-5_2}
\showDOI{\tempurl}


\bibitem[Messarra et~al\mbox{.}(2024)]%
        {IA2024Vanishingculture}
\bibfield{editor}{\bibinfo{person}{Luca Messarra}, \bibinfo{person}{Chris Freeland}, {and} \bibinfo{person}{Juliya Ziskina}} (Eds.). \bibinfo{year}{2024}\natexlab{}.
\newblock \bibinfo{booktitle}{\emph{Vanishing Culture: A Report on Our Fragile Cultural Record}}.
\newblock \bibinfo{publisher}{Internet Archive}.
\newblock
\urldef\tempurl%
\url{https://archive.org/details/vanishing-culture-report}
\showURL{%
\tempurl}


\bibitem[Nelson(2021)]%
        {nypwblog2021}
\bibfield{author}{\bibinfo{person}{Michael~L. Nelson}.} \bibinfo{year}{2021}\natexlab{}.
\newblock \bibinfo{title}{{Not Your Parents’ Web: The Scope and Archiving of the Modern Web}}.
\newblock \bibinfo{howpublished}{\url{https://ws-dl.blogspot.com/2021/10/2021-10-20-not-your-parents-web-scope.html}}.
\newblock


\bibitem[Nelson and {Van de Sompel}(2019)]%
        {memento:sagebook}
\bibfield{author}{\bibinfo{person}{Michael~L. Nelson} {and} \bibinfo{person}{Herbert {Van de Sompel}}.} \bibinfo{year}{2019}\natexlab{}.
\newblock \showarticletitle{Adding the Dimension of Time to {HTTP}}.
\newblock In \bibinfo{booktitle}{\emph{SAGE Handbook of Web History}}. \bibinfo{publisher}{SAGE Publishing}.
\newblock
\showISBNx{9781473980051}


\bibitem[Ntoulas et~al\mbox{.}(2004)]%
        {ntoulas2004whats}
\bibfield{author}{\bibinfo{person}{Alexandros Ntoulas}, \bibinfo{person}{Junghoo Cho}, {and} \bibinfo{person}{Christopher Olston}.} \bibinfo{year}{2004}\natexlab{}.
\newblock \showarticletitle{What\textquotesingle s New on the Web? The Evolution of the Web from a Search Engine Perspective}. In \bibinfo{booktitle}{\emph{Proceedings of the 13th International World Wide Web Conference (WWW)}}. \bibinfo{publisher}{ACM}, \bibinfo{address}{New York, NY, USA}, \bibinfo{pages}{1--12}.
\newblock
\urldef\tempurl%
\url{https://doi.org/10.1145/988672.988674}
\showDOI{\tempurl}


\bibitem[Pareto(1896)]%
        {pareto1896cours}
\bibfield{author}{\bibinfo{person}{V. Pareto}.} \bibinfo{year}{1896}\natexlab{}.
\newblock \bibinfo{booktitle}{\emph{Cours d'Economie Politique Professe a l'Universite de Lausanne}}.
\newblock Number v. 1. \bibinfo{publisher}{F. Rouge}.
\newblock
\showISBNx{9780608373911}
\showLCCN{10000049}
\urldef\tempurl%
\url{https://books.google.com/books?id=KjnhnQAACAAJ}
\showURL{%
\tempurl}


\bibitem[SalahEldeen and Nelson(2012)]%
        {Hany2012revolution}
\bibfield{author}{\bibinfo{person}{Hany~M. SalahEldeen} {and} \bibinfo{person}{Michael~L. Nelson}.} \bibinfo{year}{2012}\natexlab{}.
\newblock \showarticletitle{Losing My Revolution: How Many Resources Shared on Social Media Have Been Lost?}. In \bibinfo{booktitle}{\emph{Proceedings of the 16th International Conference on Theory and Practice of Digital Libraries (TPDL)}} \emph{(\bibinfo{series}{TPDL '12})}. Springer, \bibinfo{pages}{125--137}.
\newblock
\urldef\tempurl%
\url{https://doi.org/10.1007/978-3-642-33290-6_13}
\showDOI{\tempurl}


\bibitem[SalahEldeen and Nelson(2013)]%
        {Hany2013Carbondate}
\bibfield{author}{\bibinfo{person}{Hany~M. SalahEldeen} {and} \bibinfo{person}{Michael~L. Nelson}.} \bibinfo{year}{2013}\natexlab{}.
\newblock \showarticletitle{Carbon dating the web: estimating the age of web resources}. In \bibinfo{booktitle}{\emph{Proceedings of the 22nd International Conference on World Wide Web}} (Rio de Janeiro, Brazil) \emph{(\bibinfo{series}{WWW '13 Companion})}. \bibinfo{publisher}{Association for Computing Machinery}, \bibinfo{address}{New York, NY, USA}, \bibinfo{pages}{1075–1082}.
\newblock
\showISBNx{9781450320382}
\urldef\tempurl%
\url{https://doi.org/10.1145/2487788.2488121}
\showDOI{\tempurl}


\bibitem[Siddique and Alam(2019)]%
        {TweetedAt:NaumanSawood}
\bibfield{author}{\bibinfo{person}{Mohammed~Nauman Siddique} {and} \bibinfo{person}{Sawood Alam}.} \bibinfo{year}{2019}\natexlab{}.
\newblock \bibinfo{title}{{TweetedAt: Finding Tweet Timestamps for Pre and Post Snowflake Tweet IDs}}.
\newblock \bibinfo{howpublished}{\url{https://ws-dl.blogspot.com/2019/08/2019-08-03-tweetedat-finding-tweet.html}}.
\newblock


\bibitem[Swartz(2013)]%
        {aaron2013ZipNum}
\bibfield{author}{\bibinfo{person}{Aaron Swartz}.} \bibinfo{year}{2013}\natexlab{}.
\newblock \bibinfo{title}{Zipnum and CDX Cluster Merging}.
\newblock \bibinfo{howpublished}{\url{https://web.archive.org/web/20160804001009/http://aaron.blog.archive.org/2013/05/28/zipnum-and-cdx-cluster-merging/}}.
\newblock


\bibitem[{Van de Sompel} et~al\mbox{.}(2009)]%
        {nelson:memento:tr}
\bibfield{author}{\bibinfo{person}{Herbert {Van de Sompel}}, \bibinfo{person}{Michael~L. Nelson}, \bibinfo{person}{Robert Sanderson}, \bibinfo{person}{Lyudmila~L. Balakireva}, \bibinfo{person}{Scott Ainsworth}, {and} \bibinfo{person}{Harihar Shankar}.} \bibinfo{year}{2009}\natexlab{}.
\newblock \bibinfo{booktitle}{\emph{{Memento: Time Travel for the Web}}}.
\newblock \bibinfo{type}{{T}echnical {R}eport} arXiv:0911.1112. \bibinfo{institution}{arXiv}.
\newblock


\bibitem[Vitter(1985)]%
        {vitter1985random}
\bibfield{author}{\bibinfo{person}{Jeffrey~S. Vitter}.} \bibinfo{year}{1985}\natexlab{}.
\newblock \showarticletitle{Random Sampling with a Reservoir}.
\newblock \bibinfo{journal}{\emph{ACM Trans. Math. Software}} \bibinfo{volume}{11}, \bibinfo{number}{1} (\bibinfo{year}{1985}), \bibinfo{pages}{37--57}.
\newblock
\urldef\tempurl%
\url{https://doi.org/10.1145/3147.3165}
\showDOI{\tempurl}


\bibitem[Vlassenroot et~al\mbox{.}(2019)]%
        {vlassenroot2019web}
\bibfield{author}{\bibinfo{person}{Eveline Vlassenroot}, \bibinfo{person}{Sally Chambers}, \bibinfo{person}{Emmanuel Di~Pretoro}, \bibinfo{person}{Friedel Geeraert}, \bibinfo{person}{Gerald Haesendonck}, \bibinfo{person}{Alejandra Michel}, {and} \bibinfo{person}{Peter Mechant}.} \bibinfo{year}{2019}\natexlab{}.
\newblock \showarticletitle{Web archives as a data resource for digital scholars}.
\newblock \bibinfo{journal}{\emph{International Journal of Digital Humanities}}  \bibinfo{volume}{1} (\bibinfo{year}{2019}), \bibinfo{pages}{85--111}.
\newblock
\urldef\tempurl%
\url{https://doi.org/10.1007/s42803-019-00007-7}
\showDOI{\tempurl}


\bibitem[Webster(2021)]%
        {Webster2021Late90sWeb}
\bibfield{author}{\bibinfo{person}{Peter Webster}.} \bibinfo{year}{2021}\natexlab{}.
\newblock \showarticletitle{Digital Archaeology in the Web of Links: Reconstructing a Late-1990s Web Sphere}.
\newblock In \bibinfo{booktitle}{\emph{The Past Web: Exploring Web Archives}}. \bibinfo{publisher}{Springer International Publishing}, \bibinfo{address}{Cham}, \bibinfo{pages}{155--164}.
\newblock
\showISBNx{978-3-030-63291-5}
\urldef\tempurl%
\url{https://doi.org/10.1007/978-3-030-63291-5_12}
\showDOI{\tempurl}


\bibitem[Weigle(2024)]%
        {nypwfindingsblog2024}
\bibfield{author}{\bibinfo{person}{Michele~C. Weigle}.} \bibinfo{year}{2024}\natexlab{}.
\newblock \bibinfo{title}{Some URLs are Immortal, Most are Not}.
\newblock \bibinfo{howpublished}{\url{https://ws-dl.blogspot.com/2024/09/2024-09-20-some-urls-are-immortal-most.html}}.
\newblock


\bibitem[Weigle et~al\mbox{.}(2024)]%
        {weigle2024righthtml}
\bibfield{author}{\bibinfo{person}{Michele~C. Weigle}, \bibinfo{person}{Michael~L. Nelson}, \bibinfo{person}{Sawood Alam}, {and} \bibinfo{person}{Mark Graham}.} \bibinfo{year}{2024}\natexlab{}.
\newblock \showarticletitle{Right HTML, Wrong JSON: Challenges in Replaying Archived Webpages Built with Client-Side Rendering}. In \bibinfo{booktitle}{\emph{Proceedings of the 2023 ACM/IEEE Joint Conference on Digital Libraries (JCDL)}} (Santa Fe, New Mexico, USA) \emph{(\bibinfo{series}{JCDL '23})}. \bibinfo{publisher}{IEEE Press}, \bibinfo{pages}{82--92}.
\newblock
\urldef\tempurl%
\url{https://doi.org/10.1109/JCDL57899.2023.00022}
\showDOI{\tempurl}


\bibitem[{WS-DL Research Group}(2024)]%
        {nypw_githubrepo}
\bibfield{author}{\bibinfo{person}{{WS-DL Research Group}}.} \bibinfo{year}{2024}\natexlab{}.
\newblock \bibinfo{title}{{Not Your Parents Web Project}}.
\newblock \bibinfo{howpublished}{\url{https://github.com/oduwsdl/nypw}}.
\newblock


\bibitem[Wu and Davison(2005)]%
        {Wu2005cloaking}
\bibfield{author}{\bibinfo{person}{Baoning Wu} {and} \bibinfo{person}{Brian~D. Davison}.} \bibinfo{year}{2005}\natexlab{}.
\newblock \showarticletitle{Cloaking and Redirection: {A} Preliminary Study}. In \bibinfo{booktitle}{\emph{AIRWeb 2005, First International Workshop on Adversarial Information Retrieval on the Web, co-located with the {WWW} conference, Chiba, Japan, May 2005}}. \bibinfo{pages}{7--16}.
\newblock
\urldef\tempurl%
\url{http://airweb.cse.lehigh.edu/2005/wu.pdf}
\showURL{%
\tempurl}


\bibitem[Zittrain et~al\mbox{.}(2021)]%
        {zittrain2021paper}
\bibfield{author}{\bibinfo{person}{Jonathan~L. Zittrain}, \bibinfo{person}{John Bowers}, {and} \bibinfo{person}{Clare Stanton}.} \bibinfo{year}{2021}\natexlab{}.
\newblock \showarticletitle{The Paper of Record Meets an Ephemeral Web: An Examination of Linkrot and Content Drift within The New York Times}.
\newblock \bibinfo{journal}{\emph{SSRN Electronic Journal}} (\bibinfo{year}{2021}), \bibinfo{pages}{1--13}.
\newblock
\urldef\tempurl%
\url{https://doi.org/10.2139/ssrn.3833133}
\showDOI{\tempurl}


\end{thebibliography}

\end{document}